\documentclass[%
reprint,onecolumn,
superscriptaddress,
nofootinbib,
amsmath,amssymb,
aps,
,floatfix
]{revtex4-1}

\usepackage{dcolumn}
\usepackage{bm}

\usepackage{amsmath}
\usepackage{amssymb}
\usepackage{amsthm}
\usepackage{mathtools}
\usepackage{mathrsfs}
\usepackage{bm}
\usepackage{slashed} 
\usepackage{graphicx}
\usepackage{multirow}
\usepackage{tikz}
\usepackage[caption=false]{subfig}
\usepackage{relsize}	
\usepackage{array}
\usepackage{float}
\usepackage{color}
\usepackage{xcolor}
\usepackage{soul}
\usepackage{verbatim} 

\allowdisplaybreaks

\newcommand{\seq}{\begin{subequations}}
	\newcommand{\sen}{\end{subequations}}
\newcommand{\be}{\begin{eqnarray}}
	\newcommand{\ee}{\end{eqnarray}}

\newcommand{\q}{\textsf{\scriptsize q}}
\newcommand{\bsq}{{\boldsymbol q}^{\perp}}
\newcommand{\bsqq}{{\boldsymbol q}^{\perp 2}}

\newcommand{\bsb}{{\boldsymbol b}^{\perp}}

\newcommand{\es}{&=&}
\newcommand{\ps}{&+&}

\newcommand{\nn}{\nonumber}
\newcommand{\nnn}{\nonumber\\}

\newcommand*{\dif}{\mathop{}\!\mathrm{d}} 

\usepackage{braket}

\usepackage{booktabs}
\AtBeginDocument{
	\heavyrulewidth=.08em
	\lightrulewidth=.05em
	\cmidrulewidth=.03em
	\belowrulesep=.65ex
	\belowbottomsep=0pt
	\aboverulesep=.4ex
	\abovetopsep=0pt
	\cmidrulesep=\doublerulesep
	\cmidrulekern=.5em
	\defaultaddspace=.5em
}

\begin{document}
	\preprint{APS/123-QED}
	\title{Gravitational Form Factors and Mechanical Properties of Quarks in Protons: A Basis Light-Front Quantization Approach}

	\author{Sreeraj Nair}
	\email{sreeraj@impcas.ac.cn}
	\author{Chandan Mondal}
	\email{mondal@impcas.ac.cn}
	\author{Siqi Xu}
	\email{xsq234@impcas.ac.cn}
	\author{Xingbo Zhao}
	\email{xbzhao@impcas.ac.cn}
	\affiliation{Institute for Modern Physics, Chinese Academy of Sciences, Lanzhou-730000, China\\ 
		School of Nuclear Science and Technology, University of Chinese Academy of Sciences, Beijing 100049, China\\
		CAS Key Laboratory of High Precision Nuclear Spectroscopy, Institute of Modern Physics, Chinese Academy of Sciences, Lanzhou 730000, China}

	\author{Asmita Mukherjee }
	\email{asmita@phy.iitb.ac.in}
	\affiliation{Indian Institute of Technology Bombay, Powai, Mumbai 400076, India}
	
	\author{James P. Vary}
	\email{jvary@iastate.edu}
	\affiliation{Department of Physics and Astronomy, Iowa State University, Ames, Iowa 50011, USA}
	
	\collaboration{BLFQ Collaboration}
	
	\begin{abstract}
		We compute the gravitational form factors (GFFs) and study their applications for the description of the mechanical properties such as the pressure, shear force distributions, and the mechanical radius of the proton from its light-front wave functions (LFWFs) based on basis light-front quantization (BLFQ).  The LFWFs of the proton are given by the lowest
		eigenvector of a light-front effective Hamiltonian that incorporates a three-dimensional confining potential and
		a one-gluon exchange interaction with fixed coupling between the constituent quarks solved in the valence Fock sector. We find acceptable agreement between our BLFQ computations and the lattice QCD for the GFFs. Our $D$-term form factor also agrees well with the extracted data from the deeply virtual Compton scattering experiments at Jefferson Lab, and the results of different phenomenological models. The distributions of pressures and shear forces are similar to those from different models. 
	\end{abstract}

	\maketitle
	
	\section{Introduction }
	\label{intro}
	Nucleons are confined systems of partons (quarks and gluons) with pressure and energy inside, dictated by the strong interaction. However, understanding the color confinement and the mass
	generation of the nucleon in quantum chromodynamics
	(QCD) theory remains one of the
	central questions in modern particle and nuclear physics. As a
	composite system, various distribution functions, such as form factors, parton distribution functions (PDFs), generalized parton distributions (GPDs), transverse momentum dependent parton distributions (TMDs), etc., are often used to describe the structure of the nucleon. The form factors of the nucleon give critical information about many fundamental aspects of its structure. The charge and magnetization distributions are encoded in the electromagnetic form factors~\cite{Miller:2007uy,Carlson:2007xd}, while the mechanical properties of the nucleon, namely
	the mass, spin, pressure and shear force are well encoded in the gravitational form factors (GFFs)~\cite{Polyakov:2018zvc,Lorce:2018egm,Burkert:2018bqq,Shanahan:2018nnv}. Specifically, one can gain a deeper knowledge on how the nucleon is mechanically shaped by its partons by studying the Druck term (D-term) GFFs~\cite{Burkert:2018bqq}. The GFFs are parameterized as the matrix elements of the energy-momentum tensor (EMT) in the nucleon states. The components of the EMT provide how matter interacts to the gravitational field.
	Thus these form factors can be extracted by direct measurement of the interaction of the nucleon with a strong gravitational field such as a neutron star. However, due to weak gravitational interaction, the direct access to them experimentally is very difficult. An indirect way to extract the GFFs
	experimentally is from hard exclusive
	processes, for example, deeply virtual Compton scattering
	(DVCS) and deeply-virtual meson production (DVMP), which are sensitive to GPDs. The GFFs are defined as the second Mellin moments of the GPDs. The quark GPDs in the nucleon have been constrained in limited kinematic regions by DVCS and DVMP experiments at JLab~\cite{CLAS:2001wjj,CLAS:2007clm,CLAS:2020yqf}, COMPASS~\cite{dHose:2004usi}, HERA~\cite{H1:1999pji,H1:2001nez,ZEUS:2003pwh,ZEUS:2003pwh}, and HERMES~\cite{HERMES:2001bob}. The world datasets have been summarized in Refs.~\cite{Favart:2015umi,dHose:2016mda,Kumericki:2016ehc}.
	Ongoing investigations at COMPASS and within the $12$ GeV program at JLab will significantly improve accesses to these
	quantities.
	The first extraction of the quark D-term form factors from DVCS and the pressure distribution inside the proton was reported in Ref.~\cite{Burkert:2018bqq}. However, the DVCS is nearly insensitive to the gluon and thus, the gluon D-term form factor is rarely extracted. The gluon contributions to the nucleon D-term form factor will be accessed at the future high-luminosity electron-ion colliders (EICs)~\cite{AbdulKhalek:2021gbh,Anderle:2021wcy}.
	
	The nucleon matrix element of the EMT involves four
	GFFs, namely $A(Q^2)$, $B(Q^2)$, $D(Q^2)$, and $\bar{C}(Q^2)$, where $Q^2$ is the squared momentum transfer from the initial to  final nucleon. They encode information on the distributions of energy density, angular momentum, and internal forces in the interior of the nucleon. The GFFs $A$ and $B$ are
	linked to the mass and spin of the nucleon. The Ji’s sum
	rule~\cite{Ji:1996ek} relates them to the partonic contribution to the total angular momentum $J$. They are again related to the generators of the Poincare group, which provides constraints on them at $Q^2=0$, which facilitates accessing these
	form factors from the experimental data. The form factor $D(Q^2)$ is related to the basic mechanical properties of the nucleon such as the pressure and stress distributions~\cite{Polyakov:2018zvc,Lorce:2018egm,Burkert:2018bqq,Shanahan:2018nnv}. It contributes to the DVCS process when there is nonzero momentum transfer in the longitudinal direction. By virtue of energy momentum conservation, the form factor $\bar{C}(Q^2)$ contributes to both the quark and gluon parts with the same magnitude but with opposite signs, therefore $\sum_{i=q,g}\bar{C}_i(Q^2)=0$. This form factor records the reshuffling of forces between the quark and gluon subsystems inside the nucleon~\cite{Polyakov:2018exb}. 
	
	There has been significant progress in the theoretical determination of the nucleon GFFs, in particular through phenomenology, QCD inspired models, and lattice QCD. The GFFs of the nucleon have been studied via chiral perturbation theory~\cite{Chen:2001pva,Belitsky:2002jp,Dorati:2007bk}, chiral quark soliton model~\cite{ Schweitzer:2002nm, Wakamatsu:2006dy, Goeke:2007fq,Goeke:2007fp,Wakamatsu:2007uc}, Bag model~\cite{Neubelt:2019sou}, Skyrme model~\cite{Cebulla:2007ei,Kim:2012ts}, light-cone QCD sum rules at leading order~\cite{Anikin:2019kwi}, light-front quark-diquark model motivated by AdS/QCD~\cite{Chakrabarti:2015lba,Chakrabarti:2020kdc,Chakrabarti:2021mfd,Chakrabarti:2016mwn,Kumar:2017dbf}, dispersion relation~\cite{Pasquini:2014vua}, instanton picture~\cite{Polyakov:2018exb}, and instant and front form~\cite{Lorce:2018egm}, holographic QCD~\cite{Abidin:2009hr,Mondal:2015fok,Mamo:2022eui,Mamo:2021krl}, lattice QCD~\cite{Hagler:2003jd,Gockeler:2003jfa,LHPC:2010jcs,LHPC:2007blg,QCDSF-UKQCD:2007gdl,Deka:2013zha}, strongly coupled scalar theory~\cite{Cao:2023ohj}, etc., while the asymptotic behavior of the GFFs has been reported in Refs.~\cite{Hatta:2018sqd,Tanaka:2018nae}.
	On the other hand, the gluon contributions to the nucleon GFFs are far less well constrained and have so far only been investigated in  holographic light-front QCD framework~\cite{deTeramond:2021lxc}, dressed quark model~\cite{More:2023pcy} and lattice QCD~\cite{Alexandrou:2020sml,Shanahan:2018pib,Yang:2018bft,Yang:2018nqn,Alexandrou:2017oeh,Pefkou:2021fni}. To date, no experimental constraints on gluon GFFs of the nucleon have been achieved. However, they are accessible via photo/leptoproduction of $J/\psi$ and $\Upsilon$~\cite{Mamo:2019mka,Hatta:2018ina,Boussarie:2020vmu}; $J/\psi$ production is investigated in experiments, which are ongoing at JLab~\cite{GlueX:2019mkq}, whereas $\Upsilon$ production are proposed at the upcoming EICs~\cite{AbdulKhalek:2021gbh,Anderle:2021wcy}. 
	The quark GFFs for the nucleon have been extracted at JLab from DVCS~\cite{Burkert:2018bqq} and the constraints on the quark GFFs can be further improved via the GPDs accessible at several existing and upcoming experimental facilities, which include the PANDA experiment at the Facility for Antiproton and Ion Research (FAIR)~\cite{PANDA:2009yku}, the proposed EICs~\cite{AbdulKhalek:2021gbh,Anderle:2021wcy}, the nuclotron-based ion collider facility (NICA)~\cite{MPD:2022qhn}, the International Linear Collider (ILC), and the Japan proton accelerator complex (J-PARC)~\cite{Kumano:2022cje}.
	
	Our theoretical framework to study the nucleon properties is based on  basis light-front quantization (BLFQ)~\cite{Vary:2009gt}, which provides a  Hamiltonian formalism for solving the relativistic many-body bound state problem in quantum field theories~\cite{Zhao:2014xaa,Nair:2022evk,Wiecki:2014ola,Li:2015zda,Jia:2018ary,Lan:2019vui,Lan:2019rba,Tang:2018myz,Tang:2019gvn,Mondal:2019jdg,Xu:2021wwj,Lan:2019img,Qian:2020utg,Lan:2021wok,Zhu:2023lst,Xu:2022abw,Peng:2022lte,Kuang:2022vdy}.
	In this paper, we compute the quark GFFs of the proton and investigate their applications for the description of
	the mechanical properties, i.e., the distributions of pressures and shear forces inside the proton from its  light-front wave functions (LFWFs) based on the BLFQ with only the valence Fock sector of the proton considered. The LFWFs  feature all three active quarks' flavor, spin, and three dimensional spatial information on the same footing.  
	Our effective Hamiltonian incorporates a three-dimensional confinement potential embodying the light-front holography in the transverse direction~\cite{Brodsky:2014yha} and a complimentary longitudinal confinement~\cite{Li:2015zda}. The Hamiltonian also includes a one-gluon exchange interaction with fixed coupling to account for the spin structure~\cite{Mondal:2019jdg}. The nonperturbative solutions for the three-body LFWFs generated by the recent BLFQ study of the nucleon ~\cite{Xu:2021wwj} have been applied successfully to generate the electromagnetic and axial form factors, radii, PDFs, GPDs, TMDs and many other properties of the nucleon~\cite{Mondal:2019jdg,Xu:2021wwj,Hu:2022ctr,Liu:2022fvl,Mondal:2021wfq,GPD_SK,GPD_yiping}. Here, we extend those studies to calculate the proton GFFs and their application for the description of
	the mechanical properties.

	The rest of the paper is organized as follows. In Sec.~\ref{sec:formalism},  we briefly summarize the BLFQ framework for the nucleon. The  proton GFFs are evaluated within BLFQ and discussed in Sec.~\ref{sec:gffs}. We present the numerical results for the GFFs and explore the mechanical properties of proton, e.g., the pressures, energy density distributions, shear forces, and the mechanical radius in Sec.~\ref{sec:results}. At the end, we provide a brief summary and conclusions in Sec.~\ref{sec:summary}.

	\section{Proton wavefunction from basis light-front quantization}\label{sec:formalism}
	The LFWFs encoding the structural information of hadronic bound states are achieved as the eigenfunctions of the eigenvalue problem of the Hamiltonian, 
	$
	H_{\rm LF}\vert \Psi\rangle=M_{\rm h}^2\vert \Psi\rangle,
	$
	with $H_{\rm LF}$ and $M_{\rm h}^2$ being the light-front Hamiltonian and the mass square eigenvalue of the hadron, respectively.  The light-front effective Hamiltonian for the proton with quarks being the only explicit degree of freedom  
	is given by~\cite{Mondal:2019jdg}
	\begin{align}\label{hami}
		H_{\rm eff}=&\sum_a \frac{{\vec k}_{\perp a}^2+m_{a}^2}{x_a}+\frac{1}{2}\sum_{a\ne b}\kappa^4 \Big[x_ax_b({ \vec r}_{\perp a}-{ \vec r}_{\perp b})^2-\frac{\partial_{x_a}(x_a x_b\partial_{x_b})}{(m_{a}+m_{b})^2}\Big]
		\nonumber\\&+\frac{1}{2}\sum_{a\ne b} \frac{F_C 4\pi \alpha_s}{Q^2_{ab}} \bar{u}(k'_a,s'_a)\gamma^\mu{u}(k_a,s_a)\bar{u}(k'_b,s'_b)\gamma^\nu{u}(k_b,s_b)g_{\mu\nu}\,,
	\end{align}
	where ${\vec k}_{\perp a}$ and $x_a$ represent the relative transverse momentum and the longitudinal momentum fraction carried by quark $a$. $m_{a}$ defines the mass of the quark $a$, and $\kappa$ is the strength of the confining potential. The variable $\vec{r}_\perp={ \vec r}_{\perp a}-{ \vec r}_{\perp b}$ represents the transverse distance between two quarks. The last term in the Hamiltonian indicates the one-gluon exchange interaction with $Q^2_{ab}=-q^2=-(1/2)(k'_a-k_a)^2-(1/2)(k'_b-k_b)^2$ being the average momentum transfer squared and $F_C =-2/3$ is the color
	factor; $\alpha_s$ is the coupling constant; and $g_{\mu\nu}$ refers to the metric tensor. ${u}(k_a,s_a)$ corresponds to the spinor with momentum $k_a$ and spin $s_a$.
	
	In the BLFQ approach, the discretized plane-wave basis is conveniently adopted in the longitudinal direction, whereas we utilize the 2D-HO function  for the transverse direction~\cite{Vary:2009gt,Zhao:2014xaa}. Solving the eigenvalue equation of the Hamiltonian, Eq.~(\ref{hami}), in the chosen basis space provides the eigenvalues as squares of the system masses, and the eigenfunctions that describe the LFWFs. The lowest eigenfunction with the relevant symmetries is naturally specified as the proton state. The LFWFs of the proton are expressed in terms of the basis function as 
	\begin{align}
		\Psi^{\Lambda}_{\{x_i,\vec{k}_{i\perp},\lambda_i\}}=\sum_{\{n_i,m_i\}} \psi^{\Lambda}_{\{x_{i},n_{i},m_{i},\lambda_i\}} \prod_i \phi_{n_i,m_i}(\vec{k}_{i\perp};b) \,,\label{wavefunctions}
	\end{align}
	with $\psi^{\Lambda}_{\{x_{i},n_{i},m_{i},\lambda_i\}}=\braket{P, {\Lambda}|\{x_i,n_i,m_i,\lambda_i\}}$ representing the LFWF in the BLFQ basis obtained by diagnalizing Eq.~(\ref{hami}) numerically, where $\ket{P, {\Lambda}}$ defines the proton state with $P$ and $\Lambda$ being the momentum and the helicity of the state. The  2D-HO function we employ as the transverse basis function is given by
	\begin{align}
		\phi_{n,m}(\vec{k}_{\perp};b)
		=\frac{\sqrt{2}}{b(2\pi)^{\frac{3}{2}}}\sqrt{\frac{n!}{(n+|m|)!}}e^{-\vec{k}_{\perp}^2/(2b^2)}\left(\frac{|\vec{k}_{\perp}|}{b}\right)^{|m|}L^{|m|}_{n}\left(\frac{\vec{k}_{\perp}^2}{b^2}\right)e^{im\theta}\label{ho_eq}\,,
	\end{align}
	where $b$ defines its scale parameter; $n$ and $m$ correspond to the principal and orbital quantum
	numbers, respectively, and $L^{|m|}_{n}$ is the associated Laguerre polynomial. The transverse basis truncation is designated by the dimensionless cutoff parameter $N_{\rm max}$, such that $\sum_i (2n_i+| m_i |+1)\le N_{\rm max}$. The basis truncation $N_{\rm max}$ plays implicitly the role of the infrared (IR) and ultraviolet (UV) regulators for the LFWFs in the transverse direction, with an IR cutoff $\Lambda_{\rm IR}\approx b /\sqrt{N_{\rm max}}$ and a UV cutoff $\Lambda_{\rm UV}\approx b \sqrt{N_{\rm max}}$. In the discretized plane-wave basis, the longitudinal momentum fraction $x_i$ of the Fock particles is defined as
	$
	x_i=p_i^+/P^+=k_i/K,
	$
	with the dimensionless quantity  $K=\sum_i k_i$, where $k=\frac{1}{2}, \frac{3}{2}, \frac{5}{2}, ...$ signifies the choice of antiperiodic boundary conditions.  The longitudinal basis cutoff $K$ controls the numerical resolution and regulates the longitudinal direction. The multi-particle basis states have the total angular momentum projection
	$
	M_J=\sum_i\left(m_i+\lambda_i\right),
	$
	where $\lambda$ denotes the quark helicity.
	
	The parameters in the effective Hamiltonian are determined to reproduce the nucleon mass and its electromagnetic properties~\cite{Xu:2021wwj}. The model LFWFs have demonstrated significant effectiveness in analyzing a broad range of nucleon properties, including electromagnetic and axial form factors, radii, PDFs, quark helicity asymmetries, GPDs, TMDs, and angular momentum distributions, achieving notable success across various metrics.~\cite{Mondal:2019jdg,Xu:2021wwj,Mondal:2021wfq}.

	\section{Gravitational form factors}

	The gauge invariant symmetric form of the QCD EMT is given by \cite{Harindranath:1997kk}
	\be
	\theta^{\mu \nu} = \frac{1}{2}\overline{\psi}\ i\left[\gamma^{\mu}D^{\nu}+\gamma^{\nu}D^{\mu}\right]\psi - F^{\mu \lambda a}F_{\lambda a}^{\nu} + \frac{1}{4} g^{\mu \nu} \left( F_{\lambda \sigma a}\right)^2 - g^{\mu \nu} \overline{\psi} \left(
	i\gamma^{\lambda}D_{\lambda} -m
	\right)
	\psi\,, \label{emtqcd}
	\ee
	with $\psi$ and $A^{\mu}$ being the fermion and boson fields, respectively. $F^{\mu \nu}_a$ is the field strength tensor for non-Abelian gauge theory, which is expressed as
	\be 
	F^{\mu \nu}_a = \partial^{\mu} A^{\nu}_a - \partial^{\nu} A^{\mu}_a +  g \ f^{abc} A^{\mu}_b A^\nu_c\,,
	\ee
	where the covariant derivative $
	iD^{\mu} = i\overleftrightarrow{\partial}^\mu+gA^{\mu} $ such that   
	$\alpha (i\overleftrightarrow{\partial}^\mu)\beta
	= \frac{i}{2}\alpha\left(\partial^{\mu}\beta\right) -\frac{i}{2}\left(\partial^{\mu}\alpha\right) \beta$. In this work, we focus only on the fermionic part of the EMT given in Eq.~(\ref{emtqcd}). Note that the last term in Eq.~(\ref{emtqcd}) vanishes owing to the equation of motion and thus, we have the following fermionic contribution to the EMT:
	\be
	\theta^{\mu \nu}_\q = \frac{1}{2}\overline{\psi}
	i\left[\gamma^{\mu}D^{\nu}+\gamma^{\nu}D^{\mu}\right]\psi\,. \label{emtqcdforquark}
	\ee

	\label{sec:gffs}
	The matrix elements of local operators such as electromagnetic current and EMT have a precise representation using LFWFs of bound systems such as hadrons. The GFFs are linked to the matrix elements of the EMT, $\theta^{\mu \nu}$, whereas the second Mellin moment of the GPDs also gives the GFFs. For a spin $1/2$ composite system, the standard parameterization of the symmetric EMT $\theta^{\mu \nu}$ involving the GFFs reads~\cite{Ji:2012vj,Harindranath:2013goa}
	\be
	\langle P^{\prime},\Lambda^{\prime}|\,	\theta^{\mu\nu}_i(0)\,|P,\Lambda \rangle \es\overline{u}(P^{\prime},\Lambda^{\prime})\bigg[-B_i(Q^2)\frac{\overline{P}^{\mu}\ \overline{P}^{ \nu}}{M}+\left(A_i(Q^2)+B_i(Q^2)\right)\frac{1}{2}(\gamma^{\mu}\overline{P}^{\nu}+\gamma^{\nu}\overline{P}^{\mu})\nnn
	\ps C_i(Q^2)\frac{q^{\mu}q^{\nu}-q^2g^{\mu\nu}}{M}+\overline{C}_i(Q^2)M\ g^{\mu\nu}\bigg]u(P,\Lambda), 
	\label{emtparam}
	\ee
	where $\overline{u}(P',\Lambda^{\prime})$, $u(P,\Lambda)$  are the Dirac spinors and $\overline{P}^{\mu}=\frac{1}{2}(P^{\prime}+P)^{\mu}$ is the average four momentum of the system. $M$ is the mass of the system and $\Lambda \left( \Lambda^{\prime}\right)$ is the helicity of the initial (final) state of the system such that $(\Lambda, \Lambda^{\prime})\equiv \{\uparrow,\downarrow\}$. Here $\uparrow(\downarrow)$ represents the positive (negative) spin projection along $z-$ axis. The Lorentz index $(\mu, \nu)\ \equiv \{+,-,1,2\}$.  $Q^2 = -q^2$ is the square of the momentum transfer. We consider the symmetric Drell-Yan frame such that the longitudinal momentum transfer $q^+ = 0$ and the average transverse momentum $\overline{P}^{\perp} = 0$. 
	The form factors $A_\q(Q^2)$ and $B_\q(Q^2)$ are extracted form the $\theta^{++}_{\q}$ component of the EMT. In light front dynamics the conserved Noether current associated with the conserved 4-momentum is $\theta^{+\mu}_{\q}$~\cite{Li:2023izn}. Furthermore, the operator $\theta^{++}_{q}$ removes a quark (or antiquark) possessing momentum $k'$ (or $k$) and spin projection $\lambda$ along the $z$-axis, and then generates a quark (or antiquark) with identical spin and momentum $k'$ (or $k$)~\cite{Brodsky:2008pf}. The GFF $A_\q(Q^2)$ is analogous to the Dirac form factor since it is obtained by summing over the helicity conserving states, whereas $B_\q(Q^2)$, which is obtained by summing over the helicity flip states, is analogous to the Pauli form factor. The GFFs $C_\q(Q^2)$ and $\overline{C}_\q(Q^2)$ are extracted from the $\theta^{jk}_{\q}$ transverse component of the EMT such that $(j,k)\equiv(1,2)$. We compactly express the matrix elements of the EMT required to extract the four GFFs as
	\be \label{matrixelement}
	\mathcal{M}^{\mu \nu }_{\Lambda \Lambda^{\prime}} = \frac{1}{2}\left[\langle P'  ,\Lambda^{\prime}|\,	\theta^{\mu \nu }_\q(0)\,|P,\Lambda \rangle \right]\,,
	\ee
	where the proton state with momentum $P$ and helicity $\Lambda$ within the valence Fock sector can be written in terms of three-particle LFWFs,
	\be
	\label{protonstate}
	\mid P,\Lambda \rangle = \int \prod_{i=1}^{3} \left[
	\frac{dx_i d^2\vec{k}_{i\perp}}{\sqrt{x}16\pi^3}
	\right]  16 \pi^3 \delta\left( 1- \sum_{i=1}^{3} x_i\right)
	\delta^2\left(\sum_{i=1}^{3} \vec{k}_{i\perp}  \right)  \Psi^{\Lambda}_{\{x_i,\vec{k}_{i\perp},\lambda_i\}} \Bigl\vert \left\{ x_i P^+, \vec{k}_{i\perp} + x_i\vec{P}_{\perp},\lambda_i \right\} \Bigl\rangle\,.
	\ee
	The GFFs $A_\q(Q^2)$ and $B_\q(Q^2)$ can then be obtained using Eq.~(\ref{matrixelement}) as follows
	\be\label{rhsA}
	\mathcal{M}^{++}_{\uparrow \uparrow} + \mathcal{M}^{++}_{\downarrow \downarrow} \es 2\  (P^+)^2A_\q(Q^2)\,, \\
	\label{rhsB}
	\mathcal{M}^{++}_{\uparrow \downarrow} + \mathcal{M}^{++}_{\downarrow \uparrow} \es \frac{ i q^{(2)}}{M} \ (P^+)^2 B_\q(Q^2) \,,
	\ee
	while the GFF $C_\q(Q^2)$ is computed from the transverse components using
	\be\label{rhsC}
	\mathcal{M}^{12}_{\uparrow \uparrow} + \mathcal{M}^{12}_{\downarrow \downarrow} \es 4 C_\q(Q^2)q^{(1)}q^{(2)}.
	\ee

	The GFF $\overline{C}_\q(Q^2)$ can be computed from the non-conservation of the partial EMT~\cite{Lorce:2018egm} which gives us:
	
	\be
	q_{\mu}\mathcal{M}^{\mu \nu }_{\uparrow \downarrow } = \overline{C}_\q(Q^2) q^{\nu} M  \left( -q^{(1)} - i q^{(2)}\right). 	  
	\ee
	
	The GFFs $A_\q(Q^2)$ and $B_\q(Q^2)$ can be written in terms of the overlap of LFWFs upon substituting the proton valence Fock state as shown in Eq.~\ref{protonstate} into the matrix element of the energy-momentum tensor in Eq.~\ref{matrixelement} and we get the following expression:
	
	\be 
	\label{eqForAB_lfwf}
	A_\q(Q^2) &=& \frac{1}{2} \sum_{ \{\lambda_i \}} \int\left[\dif\mathcal{X}\dif\mathcal{P}_\perp\right] x_1\Big{\{}\Psi^{\uparrow*}_{\{x'_i,\vec{k}'_{i\perp},\lambda_i\}}
	\Psi^{\uparrow}_{\{x_i,\vec{k}_{i\perp},\lambda_i\}} + \Psi^{\downarrow*}_{\{x'_i,\vec{k}'_{i\perp},\lambda_i\}}
	\Psi^{\downarrow}_{\{x_i,\vec{k}_{i\perp},\lambda_i\}}\Big{\}}, \nn \\
	iq^{(2)}B_\q(Q^2) &=& M\sum_{ \{\lambda_i \}} \int\left[\dif\mathcal{X}\dif\mathcal{P}_\perp\right] x_1\Big{\{}\Psi^{\uparrow*}_{\{x'_i,\vec{k}'_{i\perp},\lambda_i\}}
	\Psi^{\downarrow}_{\{x_i,\vec{k}_{i\perp},\lambda_i\}} + \Psi^{\downarrow*}_{\{x'_i,\vec{k}'_{i\perp},\lambda_i\}}
	\Psi^{\uparrow}_{\{x_i,\vec{k}_{i\perp},\lambda_i\}}\Big{\}} ,
	\ee

	where the longitudinal and transverse momenta of the struck quark are $x'_1=x_1$ and ${k'}_{1\perp} ={k}_{1\perp} + (1-x_1){q}_{\perp}$ respectively. The spectator momenta are $x'_i=x_i$ and ${k'}_{i\perp} ={k}_{i\perp} + x_i{q}_{\perp}$ with $i=(2,3)$. The shorthand notation used for the  integration measure is as follows:
	\be
	\left[\dif\mathcal{X}\dif\mathcal{P}_\perp\right]\equiv \prod_{i=1}^3\left[\frac{\dif x_i \dif^2k_{i\perp}}{16\pi^3}\right]16\pi^3\delta\left(1-\sum_{i=1}^3 x_i\right)\times \delta^2\left(\sum_{i=1}^3k_{i\perp}\right).
	\ee

	Similarly the expression for $C_\q(Q^2)$ and $\overline{C}_\q(Q^2)$ can be extracted from the equations below:

	\be 
	\label{eqForCCbar_lfwf}
	q^{(1)}q^{(2)} C_\q(Q^2) &=& \frac{1}{8} \sum_{ \{\lambda_i \}} \int\left[\dif\mathcal{X}\dif\mathcal{P}_\perp\right] \mathcal{O}^1\mathcal{O}^2\Big{\{}\Psi^{\uparrow*}_{\{x'_i,\vec{k}'_{i\perp},\lambda_i\}}
	\Psi^{\uparrow}_{\{x_i,\vec{k}_{i\perp},\lambda_i\}} + \Psi^{\downarrow*}_{\{x'_i,\vec{k}'_{i\perp},\lambda_i\}}
	\Psi^{\downarrow}_{\{x_i,\vec{k}_{i\perp},\lambda_i\}}\Big{\}}, \nn \\
	\left( -q^{(1)} - i q^{(2)}\right) 	q^{(1)}\overline{C}_\q(Q^2)  &=& -\frac{1}{M } \sum_{ \{\lambda_i \}} \int\left[\dif\mathcal{X}\dif\mathcal{P}_\perp\right] \left(q^{(1)} \mathcal{O}^1\mathcal{O}^1 + q^{(2)} \mathcal{O}^2\mathcal{O}^1 \right)\Big{\{}\Psi^{\downarrow*}_{\{x'_i,\vec{k}'_{i\perp},\lambda_i\}}
	\Psi^{\uparrow}_{\{x_i,\vec{k}_{i\perp},\lambda_i\}}\Big{\}},
	\ee

	where the operator $\mathcal{O}^j = 2k_{\perp}^{(j)} + (1-x)q_{\perp}^{(j)}$. Identities relating the overlap of HO wavefunction pertaining to the calculation of Eq.~\ref{eqForCCbar_lfwf} is shown in Appendix~\ref{appendixB}. The LFWF with negative helicity ($\Lambda = \downarrow$) is obtained from the wavefunction with positive helicity ($\Lambda = \uparrow$) using the mirror parity symmetry \cite{PhysRevD.73.036007} which gives the following relation:
	
	\be
	\psi^{\downarrow}_{\{x_{i},n_{i},m_{i},\lambda_i\}}  = (-1)^{\sum_i m_i +1} \psi^{\uparrow}_{\{x_{i},n_{i},-m_{i},-\lambda_i\}}. 
	\ee

	\section{Numerical Results}
	\label{sec:results}
	In our calculations, the quark mass influences both the kinetic energy and the one-gluon exchange interaction (OGE) terms. We opted for different quark mass values in these terms due to the distinct physics they represent. Specifically, the kinetic mass term captures long-distance physics, while the OGE term encapsulates short-distance physics through the derived interaction arising from single-gluon exchanges between quarks. This effective OGE compensates for fluctuations spanning from the valence Fock sector to higher Fock sectors.
	
	Guided by the mass evolution of the renormalization group, the quark mass linked to gluon dynamics diminishes due to contributions from higher momentum scales \cite{Xu:2021wwj}. Consequently, our choice differentiates the quark masses in the kinetic energy (\(m_{q/k}\)) and OGE (\(m_{q/g}\)) terms \cite{PhysRevD.44.3857,PhysRevD.50.971,PhysRevD.58.096015}. Specifically, we ensure \(m_{q/k} > m_{q/g}\), with the selected values detailed in Table ~\ref{table1}. Besides these, the model includes two additional parameters: the confinement potential strength (\(\kappa\)) and the coupling constant (\(\alpha_{s}\)), both listed in Table ~\ref{table1}. Our results employ truncation parameters with values \(N_{\mathrm{max}} = 10\) and \(K = 16.5\). We calibrated these parameters against existing experimental data on the Dirac and Pauli form factors \cite{Xu:2021wwj}. Using this setup, we derive the proton's four gravitational form factors of the symmetric energy-momentum tensor and depict the mechanical properties, including the quark pressure and shear distributions inside the proton.
	\begin{table}[ht]
		\caption{Parameters of the model with a transverse basis truncation set at \(N_{\mathrm{max}} = 10\) and a longitudinal truncation at \(K = 16.5\).}
		\centering
		\begin{tabular}[t]{lcccc}
			\toprule\hline
			$m_{q/k}$~~ & ~~$m_{q/g}$~~ &~~$\kappa$~~ & ~~$\alpha_{s}$~~ &\\
			\hline
			$0.3 ~\mathrm{GeV}$ &~~$0.2 ~\mathrm{GeV}$~~ &~~$0.34 ~\mathrm{GeV}$~~ & ~~$1.1 \pm 0.1$~~ & \\
			\hline\hline
		\end{tabular}
		\label{table1}
	\end{table}%

	\begin{figure}[htp!]
		\centering
		\includegraphics[width=8.5cm,height=6cm,clip]{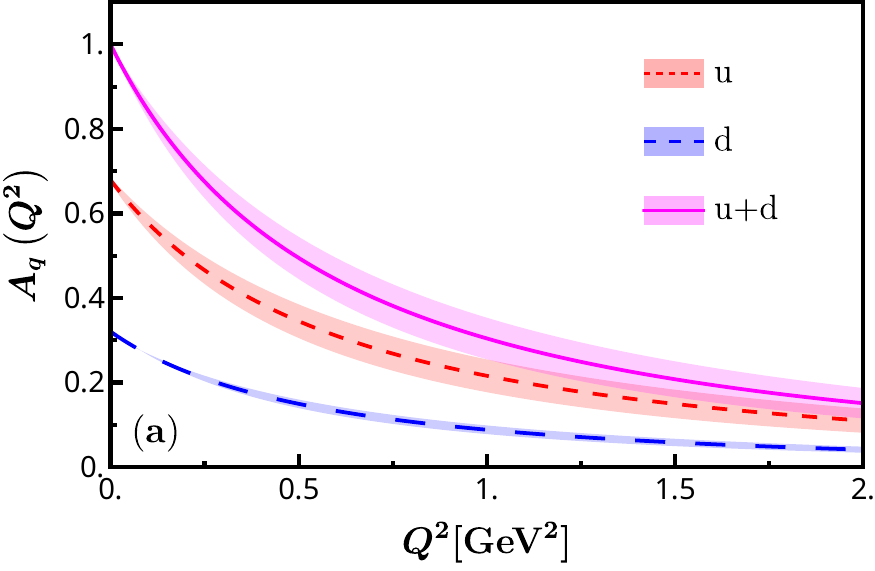}
		\includegraphics[width=8.5cm,height=6cm,clip]{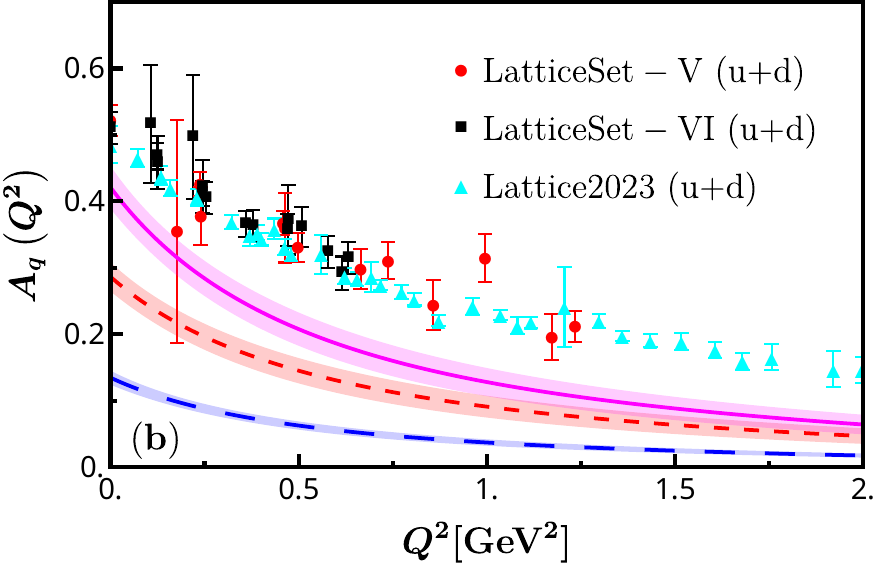}
		\includegraphics[width=8.5cm,height=6cm,clip]{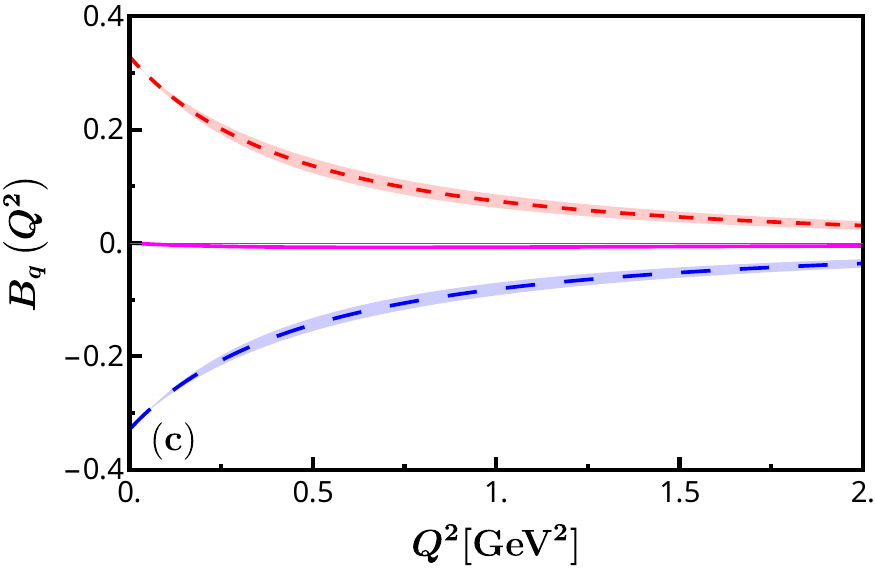}
		\includegraphics[width=8.5cm,height=6cm,clip]{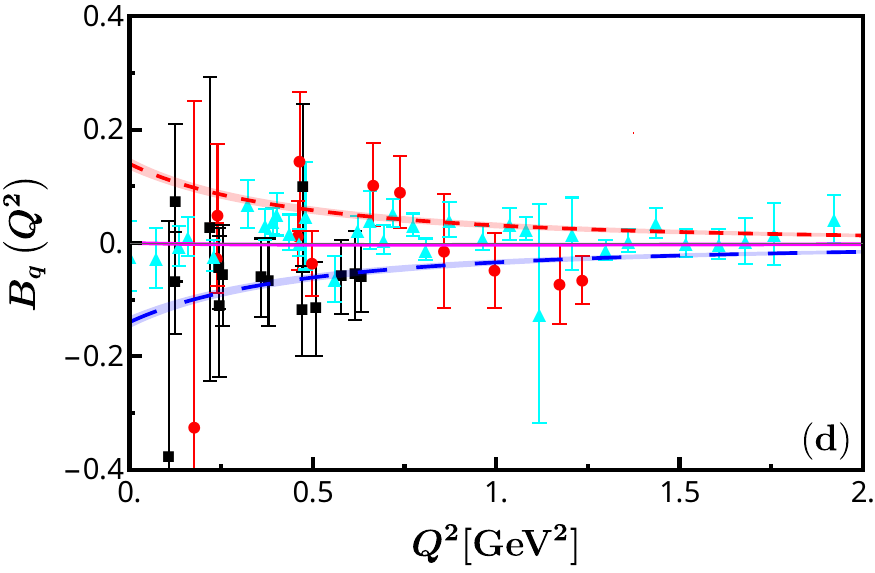}
		\caption{Plots illustrating GFFs for both  \(A_q(Q^2)\) and \(B_q(Q^2)\) at two scales, plotted against \(Q^2\). These include: (a) \(A_q(Q^2)\) at an initial scale \(\mu_0^2=0.195 \pm 0.020\) GeV\(^2\), (b) \(A_q(Q^2)\) evolved to \(\mu^2=4\) GeV\(^2\), (c) \(B_q(Q^2)\) at the same initial scale, and (d) \(B_q(Q^2)\) evolved to \(\mu^2=4\) GeV\(^2\). In the plots, solid magenta lines with shaded areas signify combined \(u\) and \(d\) quark contributions, while individual \(d\) and \(u\) quark contributions are shown with dashed blue and red lines, respectively.The shaded areas represent uncertainties: \(10\%\) in the \(\alpha_s\) coupling at the initial scale and a combined \(10\%\) in \(\mu_0^2\) and \(\alpha_s\) at the evolved scale, calculated in quadrature. The evolved GFFs are compared against lattice data at \(\mu^2=4\) GeV\(^2\), shown as red circles and black squares from Ref.~\cite{Hagler:2007xi}, and cyan triangles from Ref.~\cite{Hackett:2023rif}.\label{fig1}}
	\end{figure}

	\begin{figure}[htp!]
		\centering
		\includegraphics[width=8.5cm,height=6cm,clip]{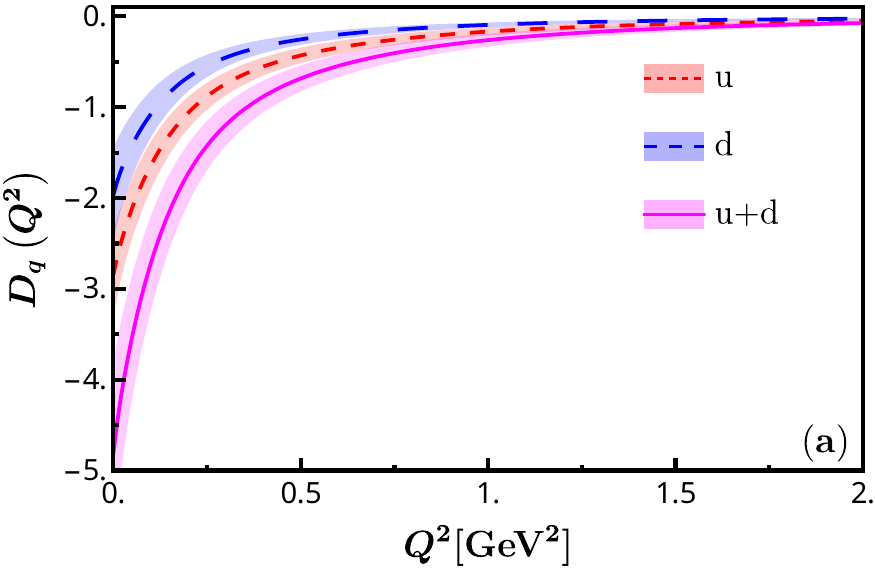}
		\includegraphics[width=8.5cm,height=6cm,clip]{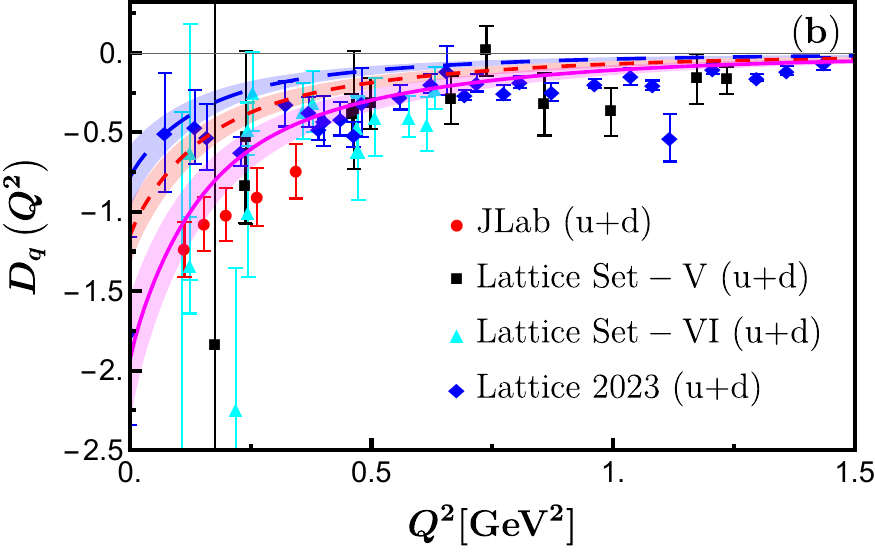}
		\includegraphics[width=8.5cm,height=6cm,clip]{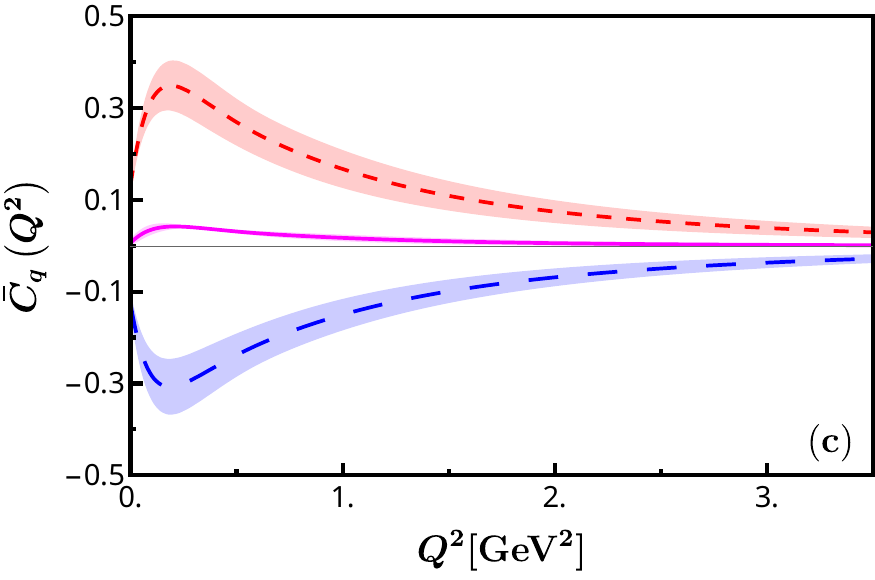}
		\includegraphics[width=8.5cm,height=6cm,clip]{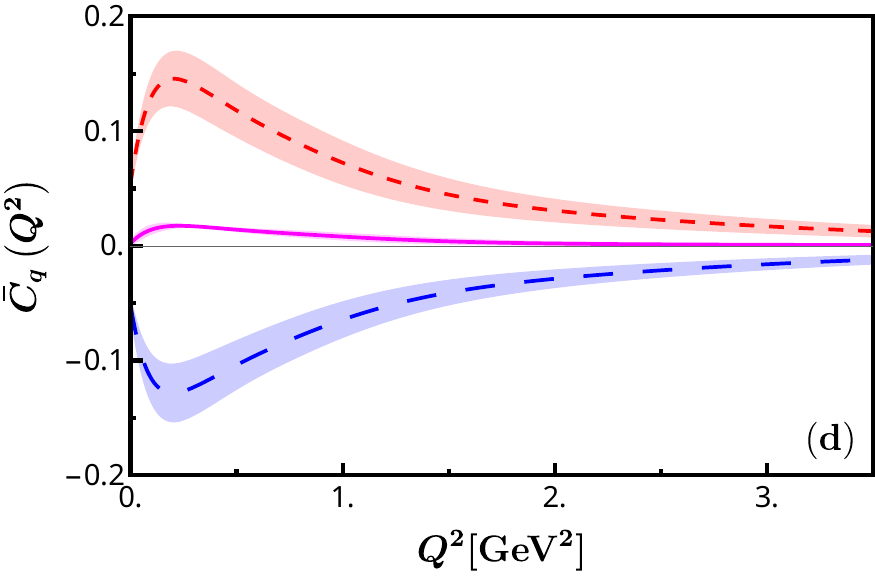}

		\caption{Plots illustrating GFFs for both \(D_q(Q^2)\) and \(\bar{C}_q(Q^2)\) across different scales: (a) \(D_q(Q^2)\) at an initial scale of \(\mu_0^2=0.195 \pm 0.020 \) GeV\(^2\), (b) \(D_q(Q^2)\) at a final evolved scale of \(\mu^2=4\) GeV\(^2\), (c) \(\bar{C}_q(Q^2)\) at the initial scale, and (d) \(\bar{C}_q(Q^2)\) at the final evolved scale, all versus \(Q^2\). The solid magenta lines with bands in each plot denote the combined contributions of the $u$ and $d$ quarks, while the dashed blue and red lines represent individual contributions from the $d$ and $u$ quarks, respectively. The bands across the plots at the initial scale signify a \(10\%\) uncertainty in the \(\alpha_s\) coupling constant whereas the bands across the evolved results are due to $10\%$ uncertainty in $\mu_0^2$ and the coupling constant $\alpha_s$. In plot (b) the red circles are the data points taken from~\cite{Shanahan:2018nnv} where they have rescaled the Jefferson Lab experimental data used in~\cite{Burkert:2018bqq} to  \(\mu^2=4\) GeV\(^2\). Lattice data are represented by cyan triangles and black squares~\cite{Hagler:2007xi}. Additionally, the evolved \(D^{u+d}_q(Q^2)\) results are compared with lattice data at \(\mu^2=4\) GeV\(^2\) as depicted by blue diamond from Ref.~\cite{Hackett:2023rif}.\label{fig2}}
	\end{figure}

	In Fig.~\ref{fig1}, we present our calculations for the GFFs \(A(Q^2)\) and \(B(Q^2)\), along with with predictions from lattice QCD~\cite{Hagler:2007xi}. The contributions from the $u$ quark and $d$ quark are individually presented for both \(A(Q^2)\) and \(B(Q^2)\) at both the initial and final scales. As the existing lattice data is at a higher scale of \(\mu^2=4\) GeV\(^2\), we've evolved our results for a consistent comparison. For the scale evolution, we employed the Dokshitzer-Gribov-Lipatov-Altarelli-Parisi (DGLAP) equations of QCD~\cite{Dokshitzer:1977sg,Gribov:1972ri,Altarelli:1977zs}, specifically utilizing the next-to-next-to-leading order (NNLO) variant.  The evolution is done at each value of $Q^2$ for the GPDs \(xH^q(x,Q^2)\) and \(xE^q(x,Q^2)\) as their moments corresponds to the GFFs \(A(Q^2)\) and \(B(Q^2)\) respectively.  The GFFs were evolved from an initial scale of $\mu_0^2=0.195 \pm 0.020$ GeV$^2$ to the lattice scale using the higher order perturbative parton evolution toolkit (HOPPET)~\cite{Salam:2008qg}. 
	The initial scale is chosen so as to match the moment of the valence quark PDFs ~\cite{Xu:2021wwj}. Our analyses indicate that at the initial scale, the values for \(A^{u+d}(0)\) and \(B^{u+d}(0)\) are 1 and 0 respectively. This is expected as a consequence of the conservation of momentum and total angular momentum. The value of \(B^{u+d}(0) = 0\)  is referred to as the vanishing of the total anomalous gravitomagnetic moment~\cite{Brodsky:2000ii}. We find that $A^u > A^d$ and $B^u$ is positive while $B^d$ is negative. Following the evolution process, the magnitudes of both functions exhibit a decline. We find that the lattice results for \(A^{u+d}(Q^2)\) significantly exceed our evolved result at \(\mu^2=4\) GeV\(^2\) especially at higher $Q^2$.

	\footnotetext[1]{The notation for the D-term used in \cite{Burkert:2018bqq}  is $d_q(Q^2) = \frac{4}{5} D_q(Q^2)$ and have $\mu^2 = 1.4~\mathrm{GeV}^2$}
	
	In Fig.~\ref{fig2}, the $D$-term form factor and the GFF \(\bar{C}(Q^2)\) are depicted. Panels (a) and (b) of Fig.~\ref{fig2} display our findings for \(D(Q^2)\) at the initial and evolved final scales, respectively. The $D$-term for both $d$ and $u$ quarks is observed to be negative, with the magnitude being larger for the $u$ quark. Following scale evolution, the characteristics of our $D$-term, as shown in Fig.~\ref{fig2} (b), correspond well with the lattice results~\cite{Hagler:2007xi,Hackett:2023rif} and JLab experimental data~\cite{Burkert:2018bqq}\protect \footnotemark. A negative $D$-term is indicative of a stable bound system, and our evolved results for the $D$-term show a reasonable alignment with the JLab data at \(\mu^2=4\) GeV\(^2\).

	Panels (c) and (d) of Fig.~\ref{fig2} present the \(\bar{C}(Q^2)\) form factor following our BLFQ framework. We observe \(\bar{C}^u(Q^2) > 0\) and \(\bar{C}^d(Q^2) < 0\), with the aggregate \(\bar{C}^{u+d}(Q^2)\) remaining near zero for most $Q^2$ values. According to the sum rule, the total \(\bar{C}^{u+d}(Q^2)\) should equal zero for all values of $Q^2$. Within our BLFQ approach, focusing on the valence Fock sector of the proton state, this condition is partially met, with $\bar{C}^{u+d}(0)$ being $0.0063$ at the initial scale and $0.0024$ at the evolved scale of \(\mu^2=4\) GeV\(^2\).
	
	In Table~\ref{compare}, we provide a detailed overview of the GFFs at \(Q^2=0\). We compare them with predictions from various phenomenological models, lattice QCD, and the current experimental data for \(D_q(0)\). For the form factors \(A_q(0)\) and \(J_q(0)\), our estimates are generally consistent, albeit slightly lower than the projections from Refs.~\cite{Hagler:2003jd,Gockeler:2003jfa,LHPC:2010jcs,Hagler:2007xi,QCDSF-UKQCD:2007gdl,Deka:2013zha,Dorati:2007bk,Lorce:2018egm} at a renormalization scale of \(\mu^2=4\) GeV\(^2\). It's noteworthy that the QCD sum rule (QCDSR) (I \& II) suggests a higher value for \(A_q(0)\), likely due to its use of a reduced scale, \(\mu^2=1\) GeV\(^2\)~\cite{Azizi:2019ytx}.
	In the $\chi$QSM and Skyrme models, only quarks and antiquarks contribute to the nucleon's angular momentum, accounting for the entirety of it. This results in the relationship \(2 J_q(0) = A_q(0) = 1\). As for the AdS/QCD models, the results are given at the model's inherent scale, with \(u\) and \(d\) quarks collectively accounting for approximately \(90\%\) of the nucleon's momentum. 
	
	Regarding the form factor \(D_q(0)\), we observe the following trends. Our estimate aligns more closely with the lattice QCD results~\cite{Gockeler:2003jfa,LHPC:2010jcs,Hagler:2007xi}, although it's slightly on the higher side. Comparatively, the predictions from QCDSR-I~\cite{Azizi:2019ytx} and QCDSR-II~\cite{Azizi:2019ytx} are in the same ballpark as our estimate, but with slight variations. On the other hand, model-based predictions such as those from the Skyrme model~\cite{Cebulla:2007ei,Kim:2012ts}, along with the \(\chi\)QSM predictions~\cite{Goeke:2007fp, Jung:2014jja, Wakamatsu:2007uc}, deviate more significantly from our results.  Interestingly, the \(\chi\)PT model~\cite{Dorati:2007bk} and IFF model~\cite{Lorce:2018egm} provide results that resonate closely with our findings. The KM15 fit~\cite{Anikin:2017fwu} and the JLab data~\cite{Burkert:2018bqq} offer values that hover around our prediction, yet show minor differences. Examining the \(\bar C^{u+d}_q(0)\) values in Table~\ref{compare}, our result has the lowest magnitude.  
	
	All GFF outcomes were fitted using a two-parameter dipole function. Comprehensive details of this function and the derived parameter values can be found in Appendix~\ref{appendixA}.
	
	\begin{table}[t]
		\caption{The GFFs for the valence quark combination at \( Q^2=0 \) are compared with predictions from other models and data from JLab. Predictions from the Skyrme and \( \chi \)QSM models take into account contributions from both quarks and gluons and are not dependent on scale. The notation \( (\mu_0 \to \mu) \) indicates the evolution from the starting scale \( \mu_0 \) to the final scale \( \mu \).
		}
		\addtolength{\tabcolsep}{2pt}
		\begin{tabular}{lccccccccc}
			\toprule\hline
			Approaches/Models ~&~ $A^{u+d}_q(0)$ ~&~ $J_q(0)=\frac{1}{2}[A^{u+d}_q(0)+B^{u+d}_q(0)]$ ~&~ $D^{u+d}(0)=4C^{u+d}(0)$~&~ $\bar C^{u+d}_q(0)$\\
			\hline\hline
			This work ($\sqrt{0.195}$ $\to$ $2$ GeV)    &  0.420 $\pm$ 0.032 & 0.210 $\pm$ 0.016 &  -1.925 $\pm$ 0.398 & 0.0024 $\pm$ 0.0034\\
			LQCD ($2$ GeV)~\cite{Hagler:2003jd}    & 0.675 & 0.34 & -&- \\
			LQCD ($2$ GeV)~\cite{Gockeler:2003jfa} & 0.547 & 0.33 & -0.80 &-\\
			LQCD ($2$ GeV)~\cite{LHPC:2010jcs}     & 0.553 & 0.238 & -1.02 &-\\
			LQCD ($2$ GeV)~\cite{Hagler:2007xi}    & 0.520 & 0.213 & -1.07&-\\
			LQCD ($2$ GeV)~\cite{QCDSF-UKQCD:2007gdl}   & 0.572 & 0.226 & - &- \\
			LQCD ($2$ GeV)~\cite{Deka:2013zha}     & 0.565 & 0.314 & - &-\\  
			$\chi$PT ($2$ GeV)~\cite{Dorati:2007bk}       & 0.538 & 0.24 & -1.44 &- \\
			IFF ($2$ GeV)~\cite{Lorce:2018egm}            & 0.55 & 0.24 &-1.28 &-0.11\\
			Asymptotic ($\infty$ GeV)~\cite{Hatta:2018sqd}            & - & 0.18 &- &-0.15\\
			QCDSR-I (1 GeV)~\cite{Azizi:2019ytx}       & 0.79 & 0.36 & -1.832 &-2.1 $\times 10^{-2}$ \\
			QCDSR-II (1 GeV)~\cite{Azizi:2019ytx}      & 0.74 & 0.30 & -1.64 &-2.5 $\times 10^{-2}$\\
			Skyrme \cite{Cebulla:2007ei}        & 1 & 0.5 & -3.584 &-\\
			Skyrme \cite{Kim:2012ts}            & 1 & 0.5 & -2.832 &-\\				
			$\chi$QSM \cite{Goeke:2007fp}       & 1 & 0.5 & -1.88 &-\\
			$\chi$QSM \cite{Jung:2014jja}       & 1 & 0.5 & -4.024 & \\
			$\chi$QSM \cite{Wakamatsu:2007uc}   & - & -& -3.88 &- \\
			AdS/QCD Model I \cite{Mondal:2015fok}       & 0.917 & 0.415& - &- \\
			AdS/QCD Model II \cite{Mondal:2015fok}       & 0.8742 & 0.392 & - &- \\
			LCSR-LO \cite{Anikin:2019kwi}       & -&- & -2.104 &- \\
			KM15 fit \cite{Anikin:2017fwu}      &-&-& -1.744  &-\\
			DR \cite{Pasquini:2014vua}          & - & - & -1.36 &-\\
			JLab data \cite{Burkert:2018bqq}    &- &-& $ -1.688 $ &-\\
			IP \cite{Polyakov:2018exb}          & -&-&-&$ 1.4 \times 10^{-2}$\\
			\hline\hline		
		\end{tabular}
		\label{compare}
	\end{table} 
	
	\subsection{Mechanical properties}\label{properties}
	
	\begin{table}[ht]
		\caption{The  mechanical properties: pressure $p_0$, energy density ${\cal E}$, and mechanical radius squared $\langle {r^2_{\text{mech}}\rangle}$ of nucleon.}
		\begin{tabular}[t]{ l c c c }
			\toprule\hline
			Approaches/Models ~&~ $p_0$  [GeV/fm$^3$]                                     ~&~ ${\cal E}$ [GeV/fm$^3$]                                  ~&~ $\langle {r^2_{\text{mech}}\rangle}$ [fm$^{2}$]\\
			\hline\hline
			This work ($\sqrt{0.195}$ GeV $\to$ $2$ GeV) &$0.47 \pm 0.10$ & $1.86 \pm 0.13$ & $0.73 \pm  0.01$\\
			QCDSR set-I (1 GeV)~\cite{Azizi:2019ytx} &$0.67$ & $1.76$ & $0.54$\\ 
			QCDSR set-II (1 GeV)~\cite{Azizi:2019ytx} & $0.62$ & $1.74$ & $0.52$\\
			Skyrme model~\cite{Cebulla:2007ei} & 0.47 & 2.28 & -\\ 
			modified Skyrme model~\cite{Kim:2012ts} & 0.26 & 1.45 & -\\
			$\chi$QSM~\cite{Goeke:2007fp} & 0.23 & 1.70 & - \\
			Soliton model~\cite{Jung:2014jja}  & 0.58 & 3.56 & -\\
			LCSM-LO~\cite{Anikin:2019kwi} & 0.84 & 0.92 & 0.54\\
			\hline\hline
		\end{tabular}
		
		\label{table2}
	\end{table}
	
	The \(D\)-term can be directly related to the pressure $p_0$ in the
	center of the nucleon and the mechanical radius squared $r^2_{\text{mech}}$ as  \cite{Polyakov:2018zvc}
	\begin{align}\label{pressure}
		p_0 &=-\frac{1}{24\pi^2 M_n} \int^{\infty}_{0} {\rm d}Q^2~ Q^3~ {D}(Q^2),\nonumber\\
		\langle r^2_{\text{mech}}\rangle&=6 D(0) \Big[ \int^{\infty}_{0} {\rm d}Q^2~ D(Q^2)\Big]^{-1}.
	\end{align}
	The energy density $\cal E$ can be obtained as follows:
	\begin{eqnarray}
		{\cal E} &=\frac{M_n}{4\pi^2} \int^{\infty}_{0} {\rm d} Q^2
		\left( A(Q^2) +
		\frac{Q^2}{4M^2_n} D(Q^2) \right),
	\end{eqnarray}
	Here, $M_n$ denotes the mass of nucleon. In Table~\ref{table2}, we list our findings alongside those from various models and approaches. For the pressure, \( p_0 \), our results are closest to those of the Skyrme model~\cite{Cebulla:2007ei} and the soliton model~\cite{Jung:2014jja}. As for the energy density, \( {\cal E} \), our value is less than that of the Skyrme model~\cite{Cebulla:2007ei} and the soliton model~\cite{Jung:2014jja} but more than most of the other predictions, including the QCDSR sets~\cite{Azizi:2019ytx} and LCSM-LO~\cite{Anikin:2019kwi}. Finally, concerning the mechanical radius squared, \( \langle {r^2_{\text{mech}}\rangle} \), our prediction surpasses the values reported in Refs.~\cite{Azizi:2019ytx,Anikin:2019kwi}.

	\subsubsection{Pressure and Shear forces}
	In our study, the \(D\)-term is derived from the transverse components of the EMT, which have connections to mechanical properties like the pressure and shear distributions~\cite{Polyakov:2002yz,Polyakov:2018zvc,Polyakov:2018exb}. A deeper understanding of these quantities can be achieved by undertaking a 2D Fourier transformation of the  \(D\)-term with respect to the transverse momentum transfer thereby  transitioning from momentum space to impact parameter space.

	The FT ($\tilde{D}(\bsb)$) of the  \(D\)-term can be expressed using the Bessel function of the zeroth order ($J_0$) as follows:
	\be
	\tilde{D}(b)
	\es  \frac{ 1}{(2\pi)^2}~\int d^2 \bsq \ e^{-i\bsq \bsb} D(q^2), \nn \\
	&=&\frac{1}{2\pi}\int_0^{\infty} d  \bsqq J_0\left( \bsq \bsb\right)D(q^2),
	\ee
	
	where, $b=|{\vec{b}_{\perp}}|$ represents the impact parameter. 
	
	The 2D pressure $p(b)$ and shear distributions $s(b)$ can thus be defined in the impact parameter space as follows ~\cite{Freese:2021czn}:
	
	\be
	\label{Prefun}
	p(b)\es\frac{1}{2M}\frac{1}{b}\frac{d}{d\ b} \left[b \frac{d}{d\ b} \tilde{D}_q(b)\right],\\
	s(b)\es-\frac{1}{M} b \frac{d}{d\ b}\left[  \frac{1}{b} \frac{d}{db} \tilde{D}_q(b)\right].
	\ee
	
	A spherical shell of radius \( b\) experiences normal and tangential forces, which are defined by the combined effects of pressure and shear~\cite{Polyakov:2018zvc}.
	
	\begin{align}
		\label{Fn-Ft}
		F_n(b)=& \left( p(b) + \frac{1}{2} s(b)\right),
		\\
		F_t(b)=& \left( p(b) - \frac{1}{2} s(b)\right).
	\end{align}
	
	\begin{figure}[htp!]
		\centering
		\includegraphics[width=8.5cm,height=6cm,clip]{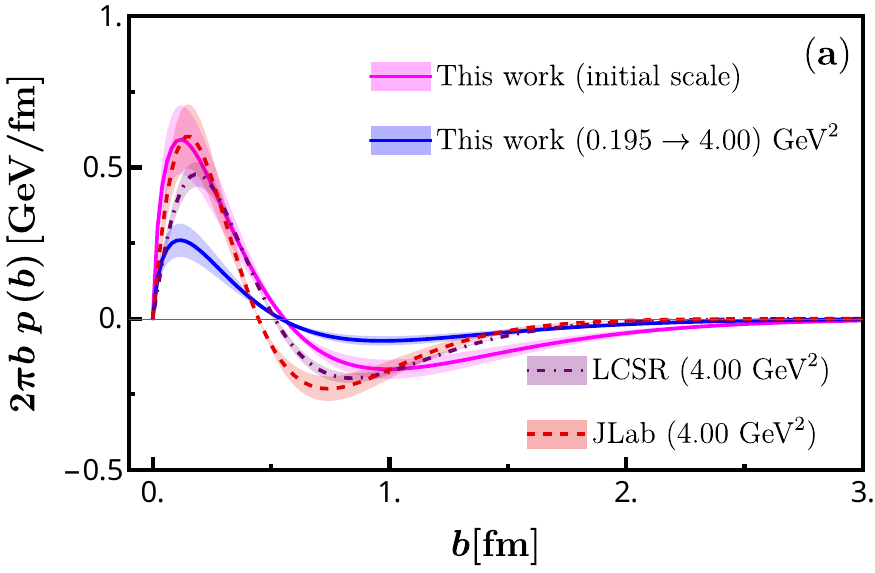}
		\includegraphics[width=8.5cm,height=6cm,clip]{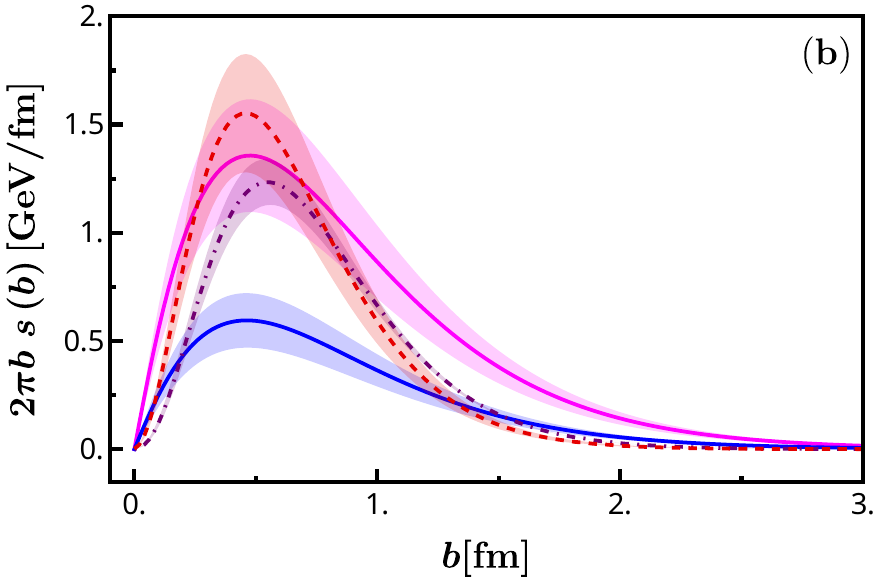}
		\caption{Plots of (a) the pressure distribution $2\pi p(b)$, and (b) the shear force distribution $2\pi s(b)$ as a function of $b$. Our results are compared with results based on light-cone sum rule (LCSR) evaluated in Ref.~\cite{Anikin:2019kwi} (purple dot-dashed lines)  and using the fitting function of $D(Q^2)$ based on JLab experimental data~\cite{Burkert:2018bqq} (red dashed lines). The solid magenta lines with magenta bands and the solid blue lines with blue bands represent the results at the initial scale $\mu_0^2=0.195 \pm 0.020$ GeV$^2$ and  the final evolved scale of $\mu^2=4.00$ GeV$^2$, respectively. The systematic uncertainties presented in the experimental extraction of the D-term~\cite{Burkert:2018bqq} were added in quadrature.}\label{fig3} 
	\end{figure}
	

	\begin{figure}[htp!]
		\centering
		\includegraphics[width=8.5cm,height=6cm,clip]{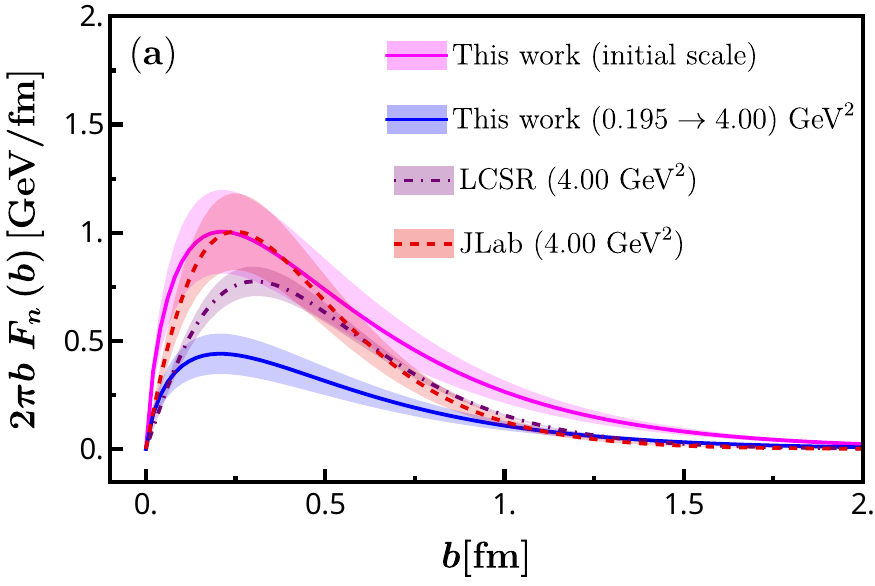}
		\includegraphics[width=8.5cm,height=6cm,clip]{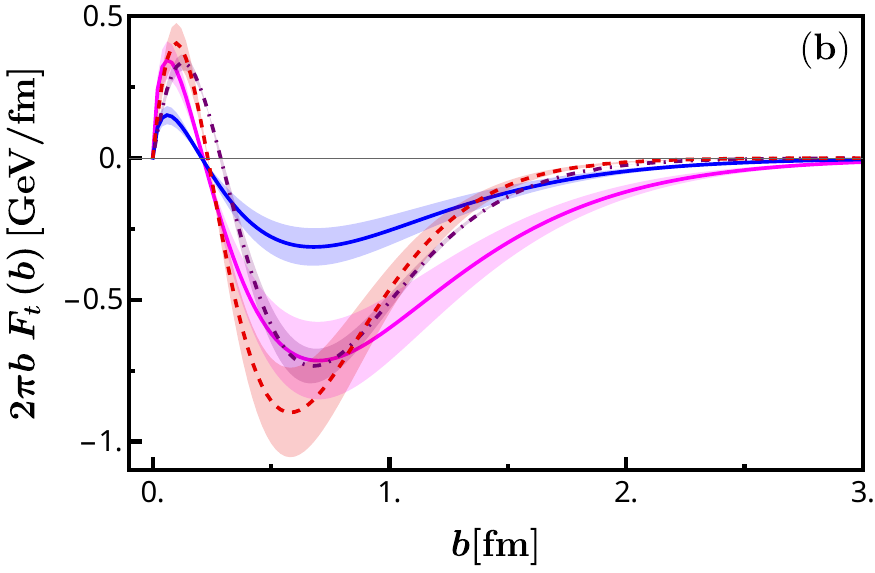}
		\caption{Plots of (a) the normal forces  $F_n$, and (b) the tangential forces $F_t$ as a function of $b$. The legends are same as mentioned in Fig.~\ref{fig3}.
		}\label{fig4} 
	\end{figure}

	In Fig.~\ref{fig3}(a), we show the distribution \( 2\pi b~p(b) \) as a function of \( b \). We compare our result with the analysis from the light-cone sum rule~\cite{Anikin:2019kwi} and the experimental data from JLab using fitting functions for \( D(Q^2) \)~\cite{Burkert:2018bqq}. The pressure distribution \( p(b) \) must follow the von Laue condition which is \(\int_{0}^{\infty} db\, b^2\, p(b) = 0\). This rule comes from the need to balance internal forces in a system \cite{Polyakov:2018zvc,Polyakov:2002yz}. This means the distribution should have a point where it changes sign to satisfy the von Laue condition. Our data shows a positive region followed by a negative region. This suggests that inside the nucleon, there are forces pushing out and forces pulling in. For our results, the point where the pressure changes sign is around \( b \approx 0.6 \) fm for both scales. For the light-cone sum rule~\cite{Anikin:2019kwi}, this point is at \( b \approx 0.51 \) fm, and for the JLab data~\cite{Burkert:2018bqq}, it's around \( b \approx 0.45 \) fm. Even though the zero-crossings in our model occur somewhat later than those in the comparative datasets, the overarching characteristics and mechanical implications of the pressure distribution align across different models and empirical findings~\cite{Shanahan:2018nnv,Anikin:2019kwi,Goeke:2007fp,Kim:2012ts,Jung:2014jja,Cebulla:2007ei}. In Fig.~\ref{fig3}(b), we display the shear force distribution \(2 \pi b s(b)\). This distribution, \(s(b)\), is associated with attributes like surface tension and surface energy, which are typically positive in stable hydrostatic systems~\cite{Polyakov:2018zvc}. In line with previous studies, our results confirm that \(s(b)\) remains positive across all \(b\) values. Furthermore, the behavior of our shear force distribution is consistent with outcomes from other methodologies~\cite{Shanahan:2018nnv,Anikin:2019kwi,Goeke:2007fp,Kim:2012ts,Jung:2014jja,Cebulla:2007ei}. It seems to be a coincidence that our initial scale results for both pressure and shear align more closely with the light-cone sum rule~\cite{Anikin:2019kwi} and the JLab data~\cite{Burkert:2018bqq} than the evolved results. Additionally, the peak position of our  results at the final scale occurs at a lower value of $b$.

	In Fig~\ref{fig4}(a) and Fig~\ref{fig4}(b), we present our results for the normal and tangential forces attributed to the valence quark combination, respectively. A salient observation is the consistently positive nature of \( F_n(b) \). On the other hand, \( F_t(b) \) showcases a dual character: a positive core, indicative of repulsive forces, and a subsequent negative domain, signifying attractive forces, with the transition (zero-crossing) located around \( b \approx 0.4 \) fm. The peak of the repulsive force occurs close to \( b \approx 0.3 \) fm, while the maximum attractive force, which plays a pivotal role in binding, is more pronounced around \( b \approx 0.6 \) fm. It's worth noting that this binding force exhibits a greater magnitude than the repulsive counterpart. Thus within the BLFQ framework, the key features of these forces exhibit good agreement with insights from the light-cone sum rule~\cite{Anikin:2019kwi}, the distributions inferred from JLab's \( D(Q^2) \) fitting function~\cite{Burkert:2018bqq}, and the predictions of the chiral quark-soliton model~\cite{Goeke:2007fp}.

	\subsubsection{The Galilean energy density and pressure distributions}
	
	Within the context of nucleonic internal structures, understanding energy density and pressure distributions offers invaluable insights. To this end, we've drawn upon the Galilean framework to study these distributions, as described in Ref.~\cite{Lorce:2018egm}. We represent the Galilean energy density $\mu(b)$, radial pressure $\sigma^r(b)$, tangential pressure $\sigma^t(b)$, isotropic pressure $\sigma(b)$, and pressure anisotropy $\Pi(b)$ through the equations below:
	
	\be 	\label{Genergy}
	\mu(b) & =& M \left[ \frac{A(b)}{2} + \overline{C}(b) + \frac{1}{4M^2}\frac{1}{b}\frac{d}{db} \left( b\frac{d}{db}\left[ \frac{B(b)}{2} - 4C(b) \right]\right) \right] ,\\
	\label{radialP}
	\sigma^r(b) & =& M  \left[ -\overline{C}(b) + \frac{1}{M^2} \frac{1}{b} \frac{d C(b)}{db} \right] , \\
	\label{tangentialP}
	\sigma^t(b) & = &M  \left[ -\overline{C}(b) + \frac{1}{M^2} \frac{d^2C(b)}{d(b)^2} \right], \\
	\label{totalP}
	\sigma(b) & = &M  \left[ -\overline{C}(b) + \frac{1}{2}\frac{1}{M^2} \frac{1}{b} \frac{d}{db}\left(b \frac{d \ C(b)}{d b} \right) \right], \\
	\label{shearlike}
	\Pi(b) & =&  M  \left[ -\frac{1}{M^2} b \frac{d}{db}\left(\frac{1}{b} \frac{dC(b)}{d b} \right) \right].
	\ee

\begin{figure}[htp!]
		\centering
		\includegraphics[width=8.5cm,height=6cm,clip]{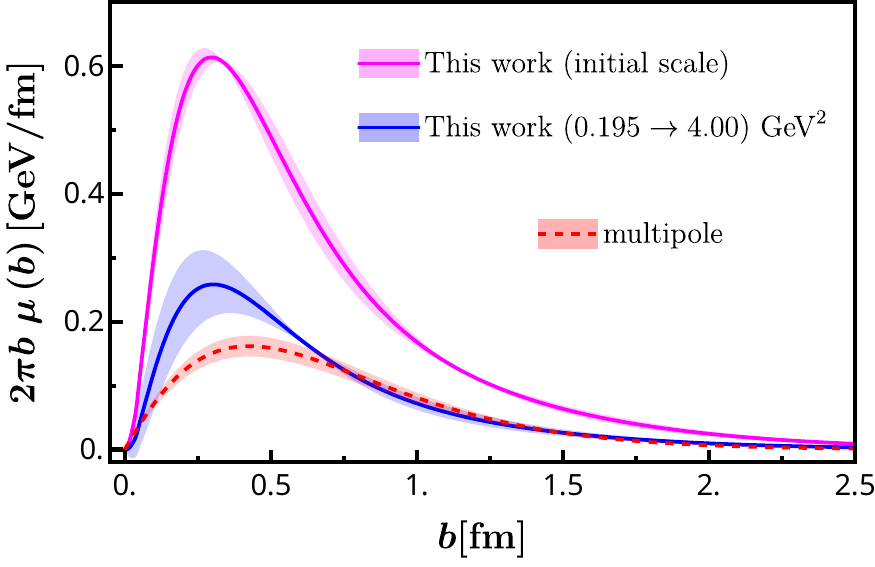}
		\caption{Plot of the two-dimensional Galilean energy density $2\pi b ~\mu_q(b)$. The solid magenta line with magenta bands and the solid blue line with blue bands represent the results at the initial scale $\mu_0^2=0.195 \pm 0.020$ GeV$^2$ and  $\mu^2=4$ GeV$^2$, respectively. Our results are compared with the results in a multipole model (red dashed line)~\cite{Lorce:2018egm} and the red band denotes a $10\%$ uncertainty in the parameter $F_q(0)$ as defined in~\cite{Lorce:2018egm}.}\label{fig5} 
	\end{figure}

	In Fig.~\ref{fig5}, we present our findings for  \(2 \pi b ~\mu \) which encompasses all four GFFs as per Eq.~\ref{Genergy}. The energy density's peak at the initial scale is discerned around \( b \approx 0.4 \) fm. However, when evolved, this peak not only diminishes in magnitude but also repositions closer, at \( b \approx 0.3 \) fm.
	
\begin{figure}[htp!]
		\centering
		\includegraphics[width=8.5cm,height=6cm,clip]{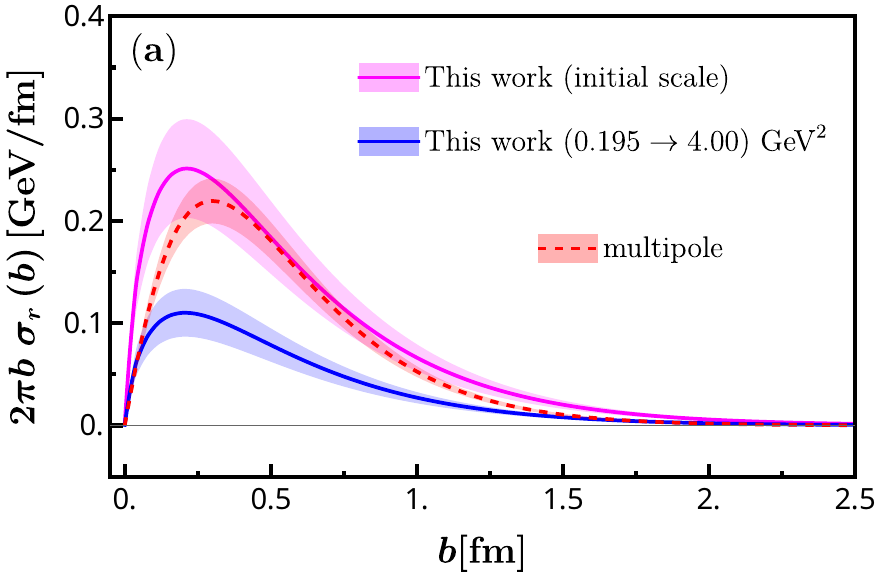}
		\includegraphics[width=8.5cm,height=6cm,clip]{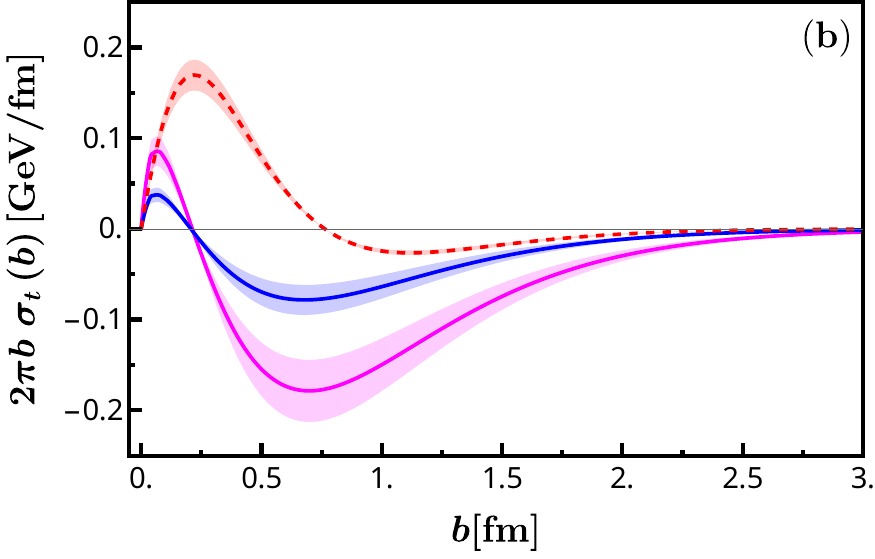}\\
		\includegraphics[width=8.5cm,height=6cm,clip]{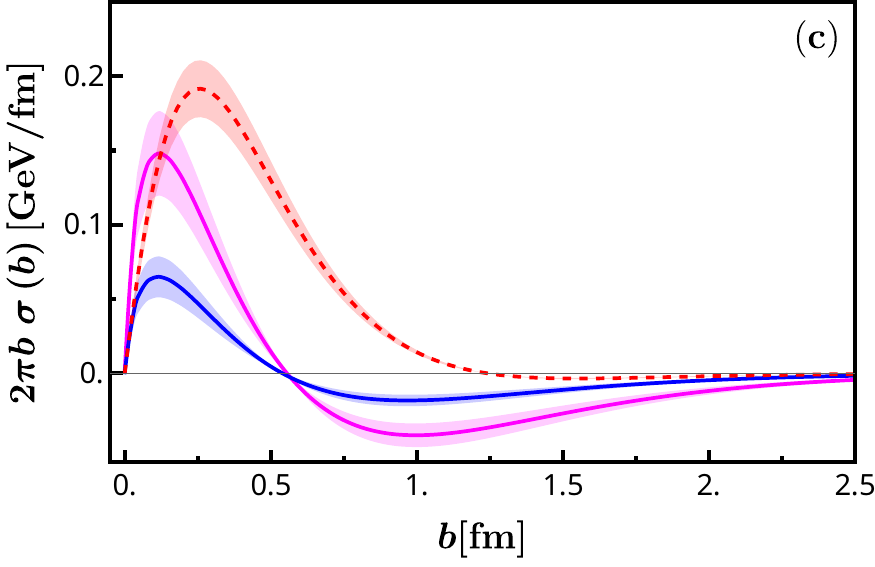}
		\includegraphics[width=8.5cm,height=6cm,clip]{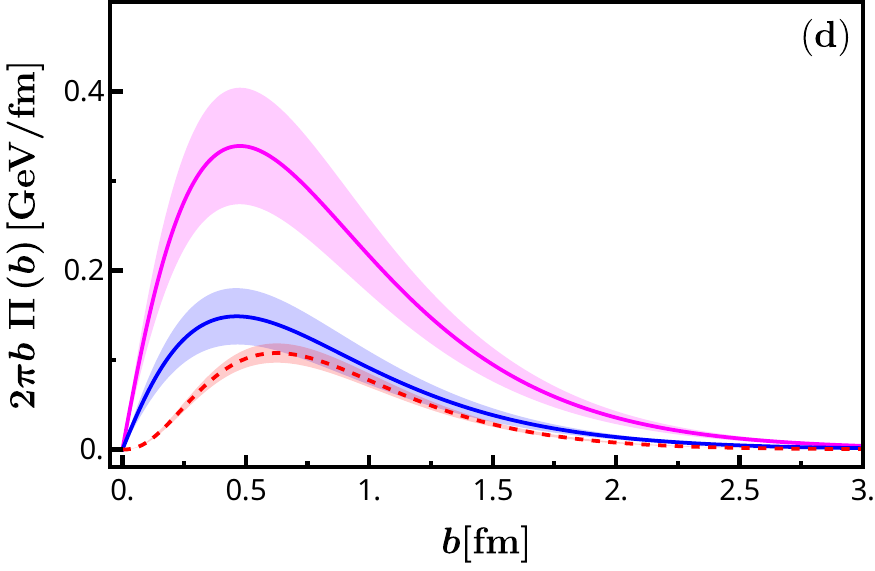}
		\caption{Plots of the two-dimensional radial pressure (a) $2\pi b ~\sigma_{r}$,  (b) tangential pressure $2\pi b ~\sigma_{t}$, (c) isotropic pressure $2\pi b ~\sigma $ and (d) the pressure anisotropy $2\pi b ~\Pi $. The legends are same as mentioned in Fig.~\ref{fig5}.
		}\label{fig6} 
	\end{figure}

	Turning to Fig.~\ref{fig6}, we elaborate on the results for the pressures defined in Eqs.~\ref{radialP}-\ref{shearlike}. The radial pressure remains consistently positive, indicating a repulsive nature. The tangential pressure, on the other hand, displays a more varied behavior: its positive (repulsive) region is located at smaller impact parameters, while the negative (attractive) region spans towards the larger values of \( b \). The isotropic pressure and the pressure anisotropy are essentially linear combinations of the radial and tangential pressures. Specifically, they are defined by the relations \( \sigma= \frac{(\sigma_r+\sigma_t)}{2} \) and \( \Pi= \sigma_r-\sigma_t \). We observe that the isotropic pressure largely mirrors the behavior of the radial pressure, though it turns slightly negative for larger \( b \) values, specifically beyond \(1.0 \) fm. The pressure anisotropy remains positive everywhere, implying that the radial pressure consistently exceeds the tangential one.
	
	Our projections for these energy and pressure distributions align well with a straightforward multipole model as proposed in Ref.~\cite{Lorce:2018egm}. An interesting coincidence  is that our results at the scale \(\mu^2=4.00\) GeV\(^2\) align more closely with the multipole model than those at the initial scale of \(\mu_0^2=0.195\) GeV\(^2\).

	\section{Conclusion}
	\label{sec:summary}
	
	In this study, we delved into the gravitational form factors (GFFs) and the mechanical attributes of quarks in the proton using the basis light-front quantization (BLFQ) theoretical approach. Utilizing an effective light-front Hamiltonian, we integrated confinement across both transverse and longitudinal planes, coupled with a one-gluon exchange interaction tailored for valence quarks, making it apt for low-resolution applications. By treating the Hamiltonian as a relativistic three-quark problem within the BLFQ framework, we deduced the nucleon light-front wave functions (LFWFs) as its eigenvectors. These LFWFs then served as the foundation to derive the quark GFFs.
	
	We evaluated the four GFFs and studied their $Q^2$ dependence at both the initial scale of  \(\mu_0^2=0.195\) GeV\(^2\) and final evolved scale of \(\mu^2=4.00\) GeV\(^2\). The $A(Q^2)$ and $B(Q^2)$ GFFs are comparable with lattice QCD results~\cite{Hagler:2007xi} although our result for the $A(Q^2)$ is slightly lower compared to  lattice QCD result. We have observed that our result for $D(Q^2)$ is in qualitative accord with the experimental data extracted from DVCS process at JLab~\cite{Burkert:2018bqq} and lattice QCD predictions~\cite{Hagler:2007xi}. We fitted our GFFs to a dipole function with two parameters as discussed in Appendix A. We have compared the values of the GFFs at $Q^2=0$ with the existing theoretical predictions and the data from JLab. Our estimates for \(A_q(0)\) and \(J_q(0)\) align reasonably with multiple references, though they tend to be slightly lower at a renormalization scale of \(\mu^2=4\) GeV\(^2\). For the \(D_q(0)\) form factor, our results are largely in sync with lattice QCD.  Our \(\bar C^{u+d}_q(0)\) value is found to be close to zero such that \(\bar C^{u}_q(0) > 0\)  and \(\bar C^{u}_q(0) < 0\) since both have the same shapes, opposite signs and nearly the same magnitudes.
	
	Using GFFs in the BLFQ framework, we assessed the proton's internal pressure, energy density, and mechanical radius, comparing with other theoretical models. The central proton pressure, \( p_0 \), aligns best with Skyrme~\cite{Cebulla:2007ei} and soliton models~\cite{Jung:2014jja}. Our energy density, \( {\cal E} \), situates between LCSM-LO~\cite{Anikin:2019kwi} and other predictions like QCDSR~\cite{Azizi:2019ytx}. Despite varying theoretical predictions on \( p_0 \) and \( \mathcal{E} \), our mechanical radius, \( \langle {r^2_{\text{mech}}\rangle} \), is notably larger than in Refs.~\cite{Azizi:2019ytx,Anikin:2019kwi}.
	
	In the BLFQ framework, we analyzed the internal pressure \(p(b)\) and shear force \(s(b)\) distributions within the proton. We observed a positive core and a negative tail for \(p(b)\), while \(s(b)\) remained consistently positive, aligning with both experimental data and other theoretical models. Additionally, the normal force \(F_n(b)\) is uniformly repulsive, whereas the tangential force \(F_t(b)\) shows repulsion at the center but attraction towards the periphery. These patterns generally concur with experimental and theoretical expectations. Moreover, our computations of the two-dimensional Galilean energy density, various pressure metrics, and anisotropy in the BLFQ framework match well qualitatively with a multipole model, especially for evolved results at higher scales.
	
	The results showcased validate the efficacy of our effective Hamiltonian method in the BLFQ framework, encouraging its use for other hadrons. Our future plans will focus on calculating the gluon GFFs within the proton system by expanding the Fock space to include an additional dynamical gluon. 
	
	\section{ACKNOWLEDGMENTS}
	
	J. P. V. acknowledges useful discussions with Yang Li. S. N. is supported by the Senior Scientist Program funded by Gansu Province, Grant No. 23JDKA0005. CM thanks the Chinese Academy of Sciences President’s International Fellowship Initiative for the support via Grants No. 2021PM0023. C. M. is also supported by new faculty start up funding by the Institute of Modern Physics, Chinese Academy of Sciences, Grant No. E129952YR0.  X. Z. is supported by new faculty startup funding by the Institute of Modern Physics, Chinese Academy of Sciences, by Key Research Program
	of Frontier Sciences, Chinese Academy of Sciences, Grant No. ZDB-SLY-7020, by the Natural Science Foundation of Gansu Province, China, Grant No. 20JR10RA067, by the Foundation for Key Talents of Gansu Province, by the Central Funds Guiding the Local Science and Technology Development of Gansu Province, Grant No. 22ZY1QA006, by international partnership program of the Chinese Academy of Sciences, Grant No. 016GJHZ2022103FN, by the National Natural Science Foundation of China under Grant No. 12375143, by National Key R\&D Program of China, Grant No. 2023YFA1606903, and by the Strategic Priority Research Program of the Chinese Academy of Sciences, Grant No. XDB34000000. A. M. would like to thank SERB MATRICS (MTR/2021/000103) for funding. J. P. V. is supported in part by the US Department of Energy, Division of Nuclear Physics,  Grant No. DE-SC0023692. A portion of the computational resources were also provided by Gansu Computing Center.

	\appendix
	
	\section{GFFs fitted with a dipole function}\label{appendixA}
	
	We fit  both the initial and final scale GFFs with the following function:
	\be\label{fit}
	f(Q^2) = \frac{a_0}{(1+a_1Q^2)^2}.
	\ee
	
	The fit parameters for the form factors \(A(Q^2)\) and \(B(Q^2)\) are listed in Table ~\ref{table_fit1}. Parameters for \(D(Q^2)\) and \(\bar{C}(Q^2)\) are provided in Table ~\ref{table_fit2}. It should be emphasized that the fit for \(\bar{C}(Q^2)\) is only reliable for \(Q^2 > 0.2\), as its fitting function does not faithfully represent \(\bar{C}(Q^2 \approx 0)\). The efficacy of the fit is evaluated through the \(\chi^2\) values presented for each GFF in Tables~\ref{table_fit1} and~\ref{table_fit2}, calculated using Eq.~\ref{chi2}. In this equation, \(N\) denotes the number of data points, \(N_p = 2\) represents the count of parameters, \(O_i\) are the calculated GFF values according to the BLFQ framework, and \(C_i\) are the values obtained from the fitting procedure. The summation in Eq.~\ref{chi2} encompasses all data points for which \(0 \leq Q^2 \leq 2\) for the GFFs, except for \(\bar{C}(Q^2)\), where the condition is \(0.2 \leq Q^2 \leq 2\).

	\be
	\chi^2 = \frac{1}{\left(N-N_p-1\right)} \sum_i \frac{\left(O_i - C_i\right)^2}{|C_i|}.
	\label{chi2}
	\ee
	
	\begin{table}[htp!]
		\centering
		\setlength{\tabcolsep}{5pt} 
		\caption{Fit parameters for \(A(Q^2)\) and \(B(Q^2)\) for \(u\) and \(d\) quark flavor. The unprimed parameters (\(a_i\)) are for the initial scale of \(\mu_0^2=0.195\) GeV\(^2\) and the primed parameters (\(a'_i\)) for the final scale of \(\mu^2=2\) GeV\(^2\). The upper and lower bounds denote the error bands around the main value. The \(\chi^2\) values indicate the goodness of fit calculated using Eq.~\ref{chi2}. For the calculation of each \(\chi^2\) value, we made use of 29 distinct points across the interval \(0 \leq Q^2 \leq 2\).}
		\label{table_fit1}
		\begin{tabular}{|c|c|c|c|c|c|c|c|}
			\hline
			GFF & Bound & \(a_0\) & \(a_1\) & \(\chi^2\)  & \(a_0'\) & \(a_1'\) & \(\chi^2\) \\ \hline
			$A^u(Q^2)$ & central & 0.6731 & 0.7762 & 0.00010378 & 0.2837 & 0.7764 & 0.00005875 \\
			$A^u(Q^2)$ & upper & 0.6787 & 0.6116 & 0.00015951 & 0.3034 & 0.6924 & 0.00011753 \\
			$A^u(Q^2)$ & lower & 0.6703 & 0.9167 & 0.00012235 & 0.2643 & 0.8863 & 6.5513 $\times 10^{-6}$ \\
			$A^d(Q^2)$ & central & 0.3188 & 0.9112 & 0.00001585 & 0.1338 & 0.9134 & 0.00001145 \\
			$A^d(Q^2)$ & upper & 0.3099 & 0.7733 & 0.00003394 & 0.1435 & 0.8714 & 0.00003479 \\
			$A^d(Q^2)$ & lower & 0.3220 & 1.0310 & 0.00002023 & 0.1241 & 0.9640 & 3.9597 $\times 10^{-7}$ \\ 
			$B^u(Q^2)$ & central & 0.3289 & 1.1102 & 2.1633 $\times 10^{-6}$ & 0.1398 & 1.1169 & 3.0792 $\times 10^{-7}$ \\
			$B^u(Q^2)$ & upper & 0.3195 & 0.9266 & 9.8522 $\times 10^{-6}$ & 0.1494 & 1.0434 & 9.6802 $\times 10^{-6}$ \\
			$B^u(Q^2)$ & lower & 0.3285 & 1.2743 & 5.0886 $\times 10^{-6}$ & 0.1303 & 1.2067 & 0.00002466 \\ 
			$B^d(Q^2)$ & central & -0.3279 & 1.0171 & 6.2102 $\times 10^{-6}$& -0.1393 & 1.0227 & 3.8685 $\times 10^{-6}$ \\
			$B^d(Q^2)$ & upper & -0.3185 & 0.8624 & 0.00003352 & -0.1296 & 1.0929 & 3.5263 $\times 10^{-6}$ \\
			$B^d(Q^2)$ & lower & -0.3274 & 1.1574 & 3.7700 $\times 10^{-6}$ & -0.1489 & 0.9649& 0.00002398 \\
			\hline
		\end{tabular}
	\end{table}

	\begin{table}[htp!]
		\centering
		\setlength{\tabcolsep}{5pt} 
		\caption{Fit parameters for \(D(Q^2)\) and \(\bar{C}(Q^2)\) for both $u$ and $d$ quark flavor. The unprimed parameters ($a_i$) are for the initial scale of $\mu_0^2=0.195$ GeV$^2$ and the primed parameter ($a'_i$) are for the final scale of $\mu^2=2$ GeV$^2$. The upper and lower bounds denote the error bands around the main value. The \(\chi^2\) values indicate the goodness of fit calculated using Eq.~\ref{chi2}. To determine the \(\chi^2\) values for the GFF \(D(Q^2)\), we utilized a set of 29 discrete data points spanning the range \(0 \leq Q^2 \leq 2\). For the GFF \(\bar{C}(Q^2)\), we calculated using 20 discrete data points within the interval \(0.2 \leq Q^2 \leq 2\).\\}
		\label{table_fit2}
		\begin{tabular}{|c|c|c|c|c|c|c|c|c|}
			\hline
			GFF & Bound & \(a_0\) & \(a_1\) & \(\chi^2\)  & \(a_0'\) & \(a_1'\) & \(\chi^2\) \\ \hline
			$D^u(Q^2)$ & central & -2.8330 & 3.1549 & 0.00008020 & -1.1460 &2.9875 & 0.00011587 \\
			$D^u(Q^2)$ & upper & -3.4741 & 3.1525 & 0.00106679 & -0.9571& 3.0757 & 0.00054817 \\
			$D^u(Q^2)$ & lower & -2.5569 & 3.2365 & 0.00034558 &-1.3347& 2.9243 & 0.00010448 \\
			$D^d(Q^2)$ & central & -1.9610 & 3.5540 & 0.00011851 &-0.7822& 3.3392 & 0.00017370\\
			$D^d(Q^2)$ & upper &-2.6775& 3.687 & 0.00020343 & -0.5869 & 3.2820 & 0.00041175 \\
			$D^d(Q^2)$ & lower & -1.6696 & 3.5211 & 0.00040762 & -0.9776& 3.3753 & 0.00010373\\ 
			$\bar{C}^u(Q^2 \geq 0.2)$ & central & 0.4924 & 0.7259 & 0.00016296 & 0.2126 & 0.7298 & 0.00023901 \\
			$\bar{C}^u(Q^2 \geq 0.2)$ & upper & 0.5856& 0.6157 & 0.00006947 & 0.2393 & 0.6349 & 0.00023820 \\
			$\bar{C}^u(Q^2 \geq 0.2)$ & lower & 0.4732 & 0.8198& 0.00013749 & 0.1892& 0.8969 & 0.00029638 \\ 
			$\bar{C}^d(Q^2 \geq 0.2)$ & central & -0.4257 & 0.6935 & 0.00011733 & -0.1839 & 0.7011 &0.00015369 \\
			$\bar{C}^d(Q^2 \geq 0.2)$ & upper & -0.5426 & 0.6545 & 0.00003208  & -0.1531 & 0.7846 & 0.00017627 \\
			$\bar{C}^d(Q^2 \geq 0.2)$ & lower & -0.3988 & 0.7485& 0.00010404 & -0.2154 & 0.6509& 0.00014935 \\
			\hline
		\end{tabular}
	\end{table}

	\section{HO integrals used for calculating $C_\q(Q^2)$ and $\overline{C}_\q(Q^2)$}\label{appendixB}

	The recurrence relation for the HO wavefunction is as follows~\cite{Wiecki:2014ola}:
	\be
	& k ~\phi_{n,m}(\vec{k}_{\perp})= 
	b \begin{cases}\sqrt{n+|m|+1} \phi_{n,m+1}(\vec{k}_{\perp})-H(n-1) \sqrt{n} \phi_{n-1,m+1}(\vec{k}_{\perp}) & ; m \geq 0 \\
		\sqrt{n+|m|} \phi_{n,m+1}(\vec{k}_{\perp})-\sqrt{n+1} \phi_{n+1,m+1}(\vec{k}_{\perp}) & ; m<0\end{cases}
	\ee

	\be
	& k^* ~\phi_{n,m}(\vec{k}_{\perp})= 
	b \begin{cases}\sqrt{n+|m|+1} \phi_{n,m-1}(\vec{k}_{\perp})-H(n-1) \sqrt{n} \phi_{n-1,m-1}(\vec{k}_{\perp}) & ; m < 0 \\
		\sqrt{n+|m|} \phi_{n,m-1}(\vec{k}_{\perp})-\sqrt{n+1} \phi_{n+1,m-1}(\vec{k}_{\perp}) & ; m\ge 0\end{cases}
	\ee
	
	where $k$ is the complex representation of $\vec{k}_{\perp}$ such that $k = \vec{k}_{\perp}^{(1)} + i \vec{k}_{\perp}^{(2)}$ and the HO wavefuntion is as defined in Eq.~\ref{ho_eq}. $H(n)$ is the unit step function. The following integrals were utilized when evaluating the GFFs $C_\q(Q^2)$ and $\overline{C}_\q(Q^2)$ and they are derived form the recurrence relation shown above.

	\be
	\mathcal{I}_1 & = 
	\int_{-\infty}^{\infty} d^2k_{\perp}~k_{\perp}^{(1)} \times \phi^{*}_{n,m}(\vec{k}_{\perp},b_1) \phi_{n',m'}(\vec{k'}_{\perp},b_2) = \frac{1}{2} \Big( \mathcal{OI}_1(x, b_1, n, m, b_2, n', m', q^{(1)}, q^{(2)}) \\ \nn &+ ~\mathcal{OI}_1^{*}(x, b_1, n, m, b_2, n', m', q^{(1)}, q^{(2)})\Big),
	\ee
	
	\be
	\mathcal{I}_2 & = 
	\int_{-\infty}^{\infty} k_{\perp}^{(2)} \times \phi^{*}_{n,m}(\vec{k}_{\perp},b_1) \phi_{n',m'}(\vec{k'}_{\perp},b_2) = -\frac{i}{2} \Big( \mathcal{OI}_1(x, b_1, n, m, b_2, n', m', q^{(1)}, q^{(2)}) \\ \nn &- ~\mathcal{OI}_1^{*}(x, b_1, n, m, b_2, n', m', q^{(1)}, q^{(2)})\Big),
	\ee
	
	\be
	\mathcal{I}_{12} & = 
	\int_{-\infty}^{\infty} k_{\perp}^{(1)} k_{\perp}^{(2)} \times \phi^{*}_{n,m}(\vec{k}_{\perp},b_1) \phi_{n',m'}(\vec{k'}_{\perp},b_2) = -\frac{i}{4} \Big( \mathcal{OI}_2(x, b_1, n, m, b_2, n', m', q^{(1)}, q^{(2)}) \\ \nn &- ~\mathcal{OI}_2^{*}(x, b_1, n, m, b_2, n', m', q^{(1)}, q^{(2)})\Big),
	\ee

	\be
	\mathcal{I}_{11} & = 
	\int_{-\infty}^{\infty} k_{\perp}^{(1)} k_{\perp}^{(1)} \times \phi^{*}_{n,m}(\vec{k}_{\perp},b_1) \phi_{n',m'}(\vec{k'}_{\perp},b_2) = \frac{1}{4} \Big( \mathcal{OI}_2(x, b_1, n, m, b_2, n', m', q^{(1)}, q^{(2)}) \\ \nn &+ ~\mathcal{OI}_2^{*}(x, b_1, n, m, b_2, n', m', q^{(1)}, q^{(2)})+ 2~\mathcal{OI}_{12}(x, b_1, n, m, b_2, n', m', q^{(1)}, q^{(2)})\Big),
	\ee

	\be
	\mathcal{I}_{22} & = 
	\int_{-\infty}^{\infty} k_{\perp}^{(2)} k_{\perp}^{(2)} \times \phi^{*}_{n,m}(\vec{k}_{\perp},b_1) \phi_{n',m'}(\vec{k'}_{\perp},b_2) = -\frac{1}{4} \Big( \mathcal{OI}_2(x, b_1, n, m, b_2, n', m', q^{(1)}, q^{(2)}) \\ \nn &+ ~\mathcal{OI}_2^{*}(x, b_1, n, m, b_2, n', m', q^{(1)}, q^{(2)})- 2~\mathcal{OI}_{12}(x, b_1, n, m, b_2, n', m', q^{(1)}, q^{(2)})\Big),
	\ee
	
	\noindent
	where $\vec{k}_{\perp}^{'} = \vec{k}_{\perp} + (1-x)q_{\perp}$ and $k_{\perp}^{(j)}$ with $j = (1,2)$ are the transverse components of $\vec{k}_{\perp}$ such that $\left(k_{\perp}^{(1)}\right)^2 + \left(k_{\perp}^{(2)}\right)^2 = \left(k_{\perp}\right)^2$. We take $b_1=b_2=b$. The functions $\mathcal{OI}$ are defined as follows:

	\begin{align*}
		&\mathcal{OI}_1(x, b_1, n, m, b_2, n', m', q^{(1)}, q^{(2)}) = \\
		& b \sqrt{x} \times \begin{cases} 
			\sqrt{n' + |m'| + 1} \times \mathcal{OL}^{b_1, b_2}_{q^{(1)}, q^{(2)}}(n, m, n', m' + 1) - H(n' - 1) \times \sqrt{n'} \times \mathcal{OL}^{b_1, b_2}_{q^{(1)}, q^{(2)}}(n, m, n' - 1, m' + 1) & \text{if } m' \geq 0 \\
			\sqrt{n' + |m'|} \times \mathcal{OL}^{b_1, b_2}_{q^{(1)}, q^{(2)}}(n, m, n', m' + 1) - \sqrt{n' + 1} \times \mathcal{OL}^{b_1, b_2}_{q^{(1)}, q^{(2)}}(n, m, n' + 1, m' + 1) & \text{otherwise}
		\end{cases}
	\end{align*}

	\begin{align*}
		&\mathcal{OI}^{*}_1(x, b_1, n, m, b_2, n', m', q^{(1)}, q^{(2)}) = \\
		& b \sqrt{x}\times \begin{cases} 
			\sqrt{n' + |m'| + 1} \times \mathcal{OL}^{b_1, b_2}_{q^{(1)}, q^{(2)}}(n, m, n', m' - 1) - H(n' - 1) \times \sqrt{n'} \times \mathcal{OL}^{b_1, b_2}_{q^{(1)}, q^{(2)}}(n, m, n' - 1, m' - 1) & \text{if } m' \leq 0 \\
			\sqrt{n' + |m'|} \times \mathcal{OL}^{b_1, b_2}_{q^{(1)}, q^{(2)}}(n, m, n', m' - 1) - \sqrt{n' + 1} \times \mathcal{OL}^{b_1, b_2}_{q^{(1)}, q^{(2)}}(n, m, n' + 1, m' - 1) & \text{otherwise}
		\end{cases}
	\end{align*}

	\begin{align}
		&\mathcal{OI}_2(x, b_1, n, m, b_2, n', m', q^{(1)}, q^{(2)}) = b^2 ~x~ f(n, m, n', m'). \\
		f(n, m, n', m') &= \begin{aligned}[t]
			&\sqrt{n' + |m'| + 1} \bigg( \sqrt{n' + |m' + 1| + 1} \mathcal{OL}^{b_1,b_2}_{q^{(1)},q^{(2)}}(n, m, n', m' + 2) \\
			&- H(n'-1) \sqrt{n'} \mathcal{OL}^{b_1,b_2}_{q^{(1)},q^{(2)}}(n, m, n'-1, m' + 2) \bigg) \\
			&- H(n'-1) \sqrt{n'} \bigg( \sqrt{n'-1 + |m' + 1| + 1} \mathcal{OL}^{b_1,b_2}_{q^{(1)},q^{(2)}}(n, m, n'-1, m' + 2) \\
			&- H(n'-2) \sqrt{n'-1} \mathcal{OL}^{b_1,b_2}_{q^{(1)},q^{(2)}}(n, m, n'-2, m' + 2) \bigg)
		\end{aligned} & \text{if } m' \geq 0 \\
		f(n, m, n', m') &= \begin{aligned}[t]
			&\sqrt{n' + |m'|} \bigg( \sqrt{n' + |m' + 1| + 1} \mathcal{OL}^{b_1,b_2}_{q^{(1)},q^{(2)}}(n, m, n', m' + 2) \\
			&- H(n'-1) \sqrt{n'} \mathcal{OL}^{b_1,b_2}_{q^{(1)},q^{(2)}}(n, m, n'-1, m' + 2) \bigg) \\
			&- \sqrt{n' + 1} \bigg( \sqrt{n' + 1 + |m' + 1| + 1} \mathcal{OL}^{b_1,b_2}_{q^{(1)},q^{(2)}}(n, m, n'+1, m' + 2) \\
			&- H(n') \sqrt{n' + 1} \mathcal{OL}^{b_1,b_2}_{q^{(1)},q^{(2)}}(n, m, n', m' + 2) \bigg)
		\end{aligned} & \text{if } m' = -1 \\
		f(n, m, n', m') &= \begin{aligned}[t]
			&\sqrt{n' + |m'|} \bigg( \sqrt{n' + |m' + 1|} \mathcal{OL}^{b_1,b_2}_{q^{(1)},q^{(2)}}(n, m, n', m' + 2) \\
			&- \sqrt{n' + 1} \mathcal{OL}^{b_1,b_2}_{q^{(1)},q^{(2)}}(n, m, n'+1, m' + 2) \bigg) \\
			&- \sqrt{n' + 1} \bigg( \sqrt{n' + 1 + |m' + 1|} \mathcal{OL}^{b_1,b_2}_{q^{(1)},q^{(2)}}(n, m, n'+1, m' + 2) \\
			&- \sqrt{n' + 2} \mathcal{OL}^{b_1,b_2}_{q^{(1)},q^{(2)}}(n, m, n'+2, m' + 2) \bigg)
		\end{aligned} & \text{otherwise}
	\end{align}

	\begin{align}
		&\mathcal{OI}_2^{*}(x, b_1, n, m, b_2, n', m', q^{(1)}, q^{(2)}) = b^2 ~x ~f^*(n, m, n', m'). \\
		f^*(n, m, n', m') &= \begin{aligned}[t]
			&\sqrt{n' + |m'| + 1} \bigg( \sqrt{n' + |m' - 1| + 1} \mathcal{OL}^{b_1,b_2}_{q^{(1)},q^{(2)}}(n, m, n', m' - 2) \\
			&- H(n'-1) \sqrt{n'} \mathcal{OL}^{b_1,b_2}_{q^{(1)},q^{(2)}}(n, m, n'-1, m' - 2) \bigg) \\
			&- H(n'-1) \sqrt{n'} \bigg( \sqrt{n' + |m' - 1|} \mathcal{OL}^{b_1,b_2}_{q^{(1)},q^{(2)}}(n, m, n'-1, m' - 2) \\
			&- H(n'-2) \sqrt{n'-1} \mathcal{OL}^{b_1,b_2}_{q^{(1)},q^{(2)}}(n, m, n'-2, m' - 2) \bigg)
		\end{aligned} & \text{if } m' \leq 0 \\
		f^*(n, m, n', m') &= \begin{aligned}[t]
			&\sqrt{n' + |m'|} \bigg( \sqrt{n' + |m' - 1| + 1} \mathcal{OL}^{b_1,b_2}_{q^{(1)},q^{(2)}}(n, m, n', m' - 2) \\
			&- H(n'-1) \sqrt{n'} \mathcal{OL}^{b_1,b_2}_{q^{(1)},q^{(2)}}(n, m, n'-1, m' - 2) \bigg) \\
			&- \sqrt{n' + 1} \bigg( \sqrt{n' + 1 + |m' - 1| + 1} \mathcal{OL}^{b_1,b_2}_{q^{(1)},q^{(2)}}(n, m, n'+1, m' - 2) \\
			&- H(n') \sqrt{n' + 1} \mathcal{OL}^{b_1,b_2}_{q^{(1)},q^{(2)}}(n, m, n', m' - 2) \bigg)
		\end{aligned} & \text{if } m' = 1 \\
		f^*(n, m, n', m') &= \begin{aligned}[t]
			&\sqrt{n' + |m'|} \bigg( \sqrt{n' + |m' - 1|} \mathcal{OL}^{b_1,b_2}_{q^{(1)},q^{(2)}}(n, m, n', m' - 2) \\
			&- \sqrt{n' + 1} \mathcal{OL}^{b_1,b_2}_{q^{(1)},q^{(2)}}(n, m, n'+1, m' - 2) \bigg) \\
			&- \sqrt{n' + 1} \bigg( \sqrt{n' + 1 + |m' - 1|} \mathcal{OL}^{b_1,b_2}_{q^{(1)},q^{(2)}}(n, m, n'+1, m' - 2) \\
			&- \sqrt{n' + 2} \mathcal{OL}^{b_1,b_2}_{q^{(1)},q^{(2)}}(n, m, n'+2, m' - 2) \bigg)
		\end{aligned} & \text{otherwise}
	\end{align}

	\begin{align}
		&\mathcal{OI}_{12}(x, b_1, n, m, b_2, n', m', q^{(1)}, q^{(2)}) = b^2~ x ~g(n, m, n', m'). \\
		g(n, m, n', m') &= \begin{aligned}[t]
			&\sqrt{n' + |m'| + 1} \bigg( \sqrt{n' + |m' + 1|} \mathcal{OL}^{b_1,b_2}_{q^{(1)},q^{(2)}}(n, m, n', m') \\
			&- \sqrt{n' + 1} \mathcal{OL}^{b_1,b_2}_{q^{(1)},q^{(2)}}(n, m, n'+1, m') \bigg) \\
			&- H(n'-1) \sqrt{n'} \bigg( \sqrt{n'-1 + |m' + 1|} \mathcal{OL}^{b_1,b_2}_{q^{(1)},q^{(2)}}(n, m, n'-1, m') \\
			&- \sqrt{n'} \mathcal{OL}^{b_1,b_2}_{q^{(1)},q^{(2)}}(n, m, n', m') \bigg)
		\end{aligned} & \text{if } m' \geq 0 \\
		g(n, m, n', m') &= \begin{aligned}[t]
			&\sqrt{n' + |m'|} \bigg( \sqrt{n' + |m' + 1| + 1} \mathcal{OL}^{b_1,b_2}_{q^{(1)},q^{(2)}}(n, m, n', m') \\
			&- H(n'-1) \sqrt{n'} \mathcal{OL}^{b_1,b_2}_{q^{(1)},q^{(2)}}(n, m, n'-1, m') \bigg) \\
			&- \sqrt{n' + 1} \bigg( \sqrt{n' + 1 + |m' + 1| + 1} \mathcal{OL}^{b_1,b_2}_{q^{(1)},q^{(2)}}(n, m, n'+1, m') \\
			&- H(n') \sqrt{n' + 1} \mathcal{OL}^{b_1,b_2}_{q^{(1)},q^{(2)}}(n, m, n', m') \bigg)
		\end{aligned} & \text{otherwise}
	\end{align}

	The function
	$\mathcal{OL}^{b_1,b_2}_{q^{(1)},q^{(2)}}$ represents the overlap of the HO wavefunction and its approximate analytic expression is shown in the following sub-section.

	\subsection{The HO overlap}

	\begin{align}
		H(n) & : \text{unit step function} \\
		\theta & = \tan^{-1}\left(\frac{b_2}{b_1}\right) \\
		l_i & = q^{(i)} \times  \sin(2\theta)\times \frac{\sqrt{b_1^2 + b_2^2}}{2} \quad (i = 1,2) \\
		l & = \sqrt{l_1^2 +l_2^2} \\
		\mathcal{C}(b_1, b_2, 	\tilde{q}) & = 
		\begin{cases} 
			1 & \text{if } b_1 = b_2 \text{ and } 	\tilde{q} = 0 \\
			\cos^{	\tilde{q}}(2\theta) & \text{otherwise}
		\end{cases}
	\end{align}
	We represent by $\mathcal{OL}^{b_1,b_2}_{q^{(1)},q^{(2)}}\left(n_1,m_1,n_2,m_2\right)$ the integral of the overlap between the harmonic oscillator wavefunction such that
	
	\be
	\mathcal{OL}^{b_1,b_2}_{q^{(1)},q^{(2)}}\left(n_1,m_1,n_2,m_2\right) = \int_{-\infty}^{\infty} d^2k_{\perp}~ \phi^{*}_{n_1,m_1}(\vec{k}_{\perp},b_1) \phi_{n_2,m_2}(\vec{k'}_{\perp},b_2),
	\ee
	
	where $\vec{k}_{\perp}^{'} = \vec{k}_{\perp} + (1-x)q_{\perp}$.

	The approximate analytic expression for the above overlap can be represented in the following form:
	
	\be
	\mathcal{OL}^{b_1,b_2}_{q^{(1)},q^{(2)}}\left(n_1,m_1,n_2,m_2\right)&=& H(n_1)H( n_2) \times \left( \frac{b_1 b_2}{b_1^2 + b_2^2} \right) \times \sqrt{2n_1! (n_1 + |m_1|)! 2n_2! (n_2 + |m_2|)!} \times (-1)^{n_1 + n_2} \times \nn \\ && \mathcal{OL}'\left(\frac{1}{b_1}, n_1, m_1, \frac{1}{b_2}, n_2, m_2, q^{(1)},q^{(2)}\right).
	\ee

	\underline{For the case $l \ne  0$ we have }
	
	\begin{equation}
		\mathcal{OL}'\left(b_1,n_1,m_1,b_2,n_2,m_2,q^{(1)},q^{(2)}\right) = \exp\left(-\frac{l^2}{2}\right) \sum_{r=r_{\text{min}}}^{r_{\text{max}}} \left( \sum_{v=v_{\text{min}}}^{v_{\text{max}}} \left( \sum_{w=w_{\text{min}}}^{w_{\text{max}}} \left( \sum_{z=z_{\text{min}}}^{z_{\text{max}}} \mathcal{EV}_0(r, v, w, z) \right) \right) \right),
	\end{equation}
	
	\be
	\mathcal{EV}_0(r, v, w, z) &=& (-1)^{\tilde{q} - u + t} \cdot l^{s + t} \cdot \frac{1}{p!} \cdot \binom{\tilde{q} + r + s + t}{\tilde{q}, r, s, t} \cdot \binom{\tilde{q}}{u} \cdot \binom{r}{v} \cdot \binom{s}{w} \cdot \binom{t}{z} \\ \nn  && \cdot \mathcal{C}(b_1, b_2, \tilde{q}) \cdot \sin^{r}(2\theta) \cdot \sin^{s}(\theta) \cdot \cos^{t}(\theta) \cdot e^{-i \cdot (s - 2w + t - 2z) \cdot  \tan^{-1}\left(l_2/l_1\right)},
	\ee
	
	\begin{equation}
		\begin{aligned}
			r_{\text{min}} &= 0, & r_{\text{max}} &= n_1 + n_2 + \frac{|m_1| - m_1}{2} + \frac{|m_2| + m_2}{2}, \\
			v_{\text{min}} &= \max\left(-n_2 - \frac{|m_2| + m_2}{2} + r, 0\right), & v_{\text{max}} &= \min\left(n_1 + \frac{|m_1| - m_1}{2}, r\right), \\
			w_{\text{min}} &= \max\left(0, r - m_1 - 2v\right), & w_{\text{max}} &= \min\left(r_{\text{max}} - r, n_1 + \frac{|m_1| - m_1}{2} - v\right), \\
			z_{\text{min}} &= 
			\begin{cases} 
				\max\left(m_2 + 2v - r, 0, r_{\text{max}} - r - w\right) & \text{if } b_1 = b_2 \\
				\max\left(m_2 + 2v - r, 0\right) & \text{otherwise}
			\end{cases}, & z_{\text{max}} &= \min\left(r_{\text{max}} - r - w, n_2 + \frac{|m_2| + m_2}{2} - r + v\right), \\
			s &= m_1 + 2v + 2w - r, & t &= -m_2 - 2v + 2z + r, \\
			\tilde{q} &= n_1 + n_2 + \frac{|m_1| - m_1}{2} + \frac{|m_2| + m_2}{2} - r - w - z, & u &= n_1 + \frac{|m_1| - m_1}{2} - v - w, \\
			p &= \tilde{q} + r + s + t.
		\end{aligned}
	\end{equation}

	\underline{For the case $l = 0$ we have }

	\be
	\mathcal{OL}'\left(b_1,n_1,m_1,b_2,n_2,m_2,q^{(1)},q^{(2)}\right) = \sum_{v=0}^{v_{\mathrm{max}}} (-1)^{ \tilde{q} - u} \times \frac{1}{p!} \times \mathcal{C}(b_1, b_2,  \tilde{q} ) \times \sin(2\theta)^r \binom{p}{ \tilde{q} } \times \binom{r}{v} \times \binom{ \tilde{q} }{u},
	\ee
	
	where
	\begin{align*}
		p &=  \tilde{q}  + r \\
		\tilde{q}  &= n_1 + n_2 - 2v \\
		r &= |m_1| + 2v \\
		u &= n_1 - v \\
		v_{\mathrm{max}}& = \min\left(\frac{n_1 + n_2}{2}, n_1\right)\\
	\end{align*}

	\bibliographystyle{elsarticle-num}
	\bibliography{references}	
\end{document}